\documentclass[1p]{elsarticle}%
\pdfoutput=1
\usepackage{hyperref}
\usepackage{amssymb}
\usepackage{latexsym}
\usepackage{epsfig}
\usepackage{dcolumn}
\usepackage{amsmath}
\usepackage{amsfonts}
\usepackage{bm}
\usepackage{color}
\usepackage{graphicx}%
\journal{Nuclear Physics B}
\begin{document}
\begin{frontmatter}
\title{Multipoint Green's functions in 1+1 dimensional integrable quantum field theories.}
\author[b]{H. M. Babujian}
\ead{babujian@yerphi.am}
\author[k]{M. Karowski}
\ead{karowski@physik.fu-berlin.de}
\ead[url]{http://users.physik.fu-berlin.de/$^\sim$karowski}
\author[t]{A.~M.~Tsvelik}
\address[b]{Yerevan Physics Institute, Alikhanian Brothers
2, Yerevan, 375036 Armenia , International Institute of Physics, Universidade
Federal do Rio Grande do Norte (UFRN), 59078-400 Natal-RN, Brazil and Simons
Center for Geometry and Physics, Stony Brook University, Stony Brook, NY
11794, USA}
\address[k]{Institut f\"{u}r Theoretische Physik, Freie Universit\"{a}t Berlin,
Arnimallee 14, 14195 Berlin, Germany}
\address[t]{Brookhaven National Laboratory, Upton, NY
11973-5000, USA}
\begin{abstract}
We calculate the multipoint Green functions in 1+1 dimensional integrable
quantum field theories. We use the crossing formula for general models and
calculate the 3 and 4 point functions taking in to account only the lower
nontrivial intermediate states contributions. Then we apply the general
results to the examples of the scaling $Z_{2}$ Ising model, sinh-Gordon model
and $Z_{3}$ scaling Potts model. We demonstrate this calculations explicitly.
The results can be applied to physical phenomena as for example to the Raman scattering.
\end{abstract}
\end{frontmatter}

\section{Introduction}

A complete set of dynamical correlation functions contains the entire
information about a given system. Unfortunately, in practice only few such
functions can be measured by available experimental techniques. Usually
experiments, such as neutron scattering measurements, probe two point
functions. However, there are exceptions and there are several experimental
techniques such as resonance Raman and resonance X-ray scattering which
measure four-point functions or even something more complicated
\cite{shvaika,DH,mukamel,cundiff}. These higher order correlation functions
carry information about the nonlinear dynamics which is especially important
and interesting in strongly correlated models. It is also an interesting
theoretical problem since such models usually require some special
non-perturbative approaches. The latter fact brings us to (1+1)-dimensional
models where such approaches are available.

The problem becomes especially interesting for massive quantum field theories
where almost nothing is known about multipoint correlation functions.
Meanwhile, as will be demonstrated in this paper, it is possible to calculate
them by using the results for matrix elements or form factors of various
operators. In the present paper we will obtain three- and four-point functions
for massive integrable models in (1+1)-dimensions. For a low particle
intermediate state approximation we apply the general results to three models.
We calculate correlation functions of the order parameter fields for the
off-critical $Z_{2}$ Ising model and the $Z_{3}$ Potts model perturbed by the
thermal operator and for the fundamental field in the sinh-Gordon model. The
models are chosen in a sequence of increasing complexity: the Ising model is
equivalent to the model of non-interacting massive Majorana fermions with a
trivial S-matrix, the sinh-Gordon model is very similar to the Ising one, but
has the simplest possible nontrivial S-matrix (a diagonal one without poles),
and the $Z_{3}$ model takes the complexity one step further having a diagonal
S-matrix with one pole on the physical sheet corresponding to a bound state of
the fundamental particles. In this article we will explore the crossing
formula \cite{Sm,BFKZ} in order to start calculation of the multipoint Green
functions or Wightman functions in 1+1 dimensional integrable quantum field
theories. It is well known that the n-particle form factor of the local field
$\phi(x)$%
\[
\langle0|\phi(0)|\theta_{1},...\theta_{n}\rangle
\]
is an analytic function of the variables $\theta_{1},..,\theta_{n}$. More
general form factors as
\[
\langle\theta_{1},...,\theta_{n}|\varphi(0)|\theta_{n+1},...,\theta
_{n+k}\rangle
\]
already are not functions but distributions or generalized functions
\cite{Sm,BFKZ,BK}. In fact the crossing formula is defining the generalized
form factors in the language of simple form factors. For example, in the case
of the 3-particle form factor we have\footnote{See (\ref{cr}).}%
\begin{multline*}
\langle\theta_{1}|\varphi(0)|\theta_{2},\theta_{3}\rangle=\langle
0|\varphi(0)|\theta_{1}+i\pi-i\epsilon,\theta_{2},\theta_{3}\rangle\\
+\delta_{\theta_{1}\theta_{2}}\langle0|\varphi(0)|\theta_{3}\rangle
+\delta_{\theta_{1}\theta_{3}}\langle0|\varphi(0)|\theta_{2}\rangle
S(\theta_{23})
\end{multline*}
The $\epsilon$-prescription and the $\delta$-functions makes the left hand
side a distribution. In more complicated cases we can define the generalized
form factors as explained in \cite{Sm,BK}. In this article we will consider 3
and 4 point Green functions. Using this definition we will evaluate the
multipoint correlators or Green functions defined as time order products of
operators:
\[
\langle0|T\varphi_{1}(x_{1})\varphi_{2}(x_{2})...\varphi_{n}(x_{n})|0\rangle
\]
We will transform these correlators into sums of products of matrix elements
inserting between the fields the identity
\[
\sum|\theta_{1},...\theta_{n}\rangle\langle\theta_{n},...\theta_{1}|\,=1
\]
and then using the crossing formula we will step by step calculate the
Wightman and Green's functions.

The results can be applied to physical phenomena as for example to Raman
scattering and nonlinear susceptibility \cite{BKT}.

\section{Green's functions}

Below in this Section we will do our calculations in the most general form
valid for all integrable models. In the next sections we will apply the
results to several concrete examples. We will concentrate on the most
difficult case of the four-point function, the calculations of the three-point
one are comparatively straightforward.

The Green's functions are time ordered n-point functions, written as a sum
over all permutations of the fields $\varphi_{i}$ and variables $x_{i}$%
\begin{equation}
\tau_{\underline{\varphi}}(\underline{x})=\langle\,0\,|\,T\varphi_{1}%
(x_{1})\,\dots\,\varphi_{n}(x_{n})|\,0\,\rangle=\sum_{\pi\in S_{n}}%
\Theta(\underline{\pi x^{0}})\,w_{\underline{\pi\varphi}}(\underline{\pi
x})\label{tau}%
\end{equation}
where $w_{\underline{\pi\varphi}}(\underline{\pi x})=\left\langle
\,0\,|\,\varphi_{\pi1}(x_{\pi1})\dots\,\varphi_{\pi n}(x_{\pi n}%
)|\,0\,\right\rangle $ is the Wightman function and \newline$\Theta
(\underline{\pi t})=\Theta(t_{\pi1}-t_{\pi2})\dots\,\Theta(t_{\pi\left(
n-1\right)  }-t_{\pi n})$. The Fourier transform is the Green's function in
momentum space%
\begin{align}
\tilde{\tau}_{\underline{\varphi}}(\underline{k})  &  =\sum_{\pi\in S_{n}}%
\int\underline{d^{2}x}e^{ix_{i}k_{i}}\Theta(\underline{\pi x^{0}%
})\,\left\langle \,0\,|\,\varphi_{\pi1}(x_{\pi1})\dots\,\varphi_{\pi n}(x_{\pi
n})|\,0\,\right\rangle \,\\
&  =\left(  2\pi\right)  ^{2}\delta^{(2)}\left(
{\textstyle\sum}
k_{i}\right)  \,\tilde{\Xi}_{\underline{\varphi}}(\underline{k})\,\label{Xi}%
\end{align}
where we have used translation invariance and split off the energy momentum
$\delta$-function defining $\tilde{\Xi}(\underline{k})$. The full Green's
function may be decomposed into the connected ones%
\[
\tilde{\tau}_{\underline{\varphi}}(\underline{k})=\sum_{\underline{k}_{1}%
\cup\dots\cup\underline{k}_{m}=\underline{k}}\tilde{\tau}_{c}(\underline
{k}_{1})\dots\tilde{\tau}_{c}(\underline{k}_{m})\,.
\]

\subsection{The Green's functions in low particle approximation}

Inserting sets of intermediate states $|\underline{p^{(j)}}\rangle
=|p_{1}^{(j)},\dots,p_{n_{j}}^{(j)}\rangle$ in (\ref{tau}) we obtain (see
\ref{a1})%
\begin{align}
  \tilde{\Xi}_{\underline{\varphi}}(\underline{k})
&  =\sum_{\pi\in S_{n}}\sum\frac{1}{\underline{n}!}\int_{\underline{p^{(1)}}%
}\dots\int_{\underline{p^{(n-1)}}}\langle\varphi_{\pi1}(0)|\underline{p^{(1)}%
}\rangle\langle\underline{p^{(1)}}|\dots\underline{|p^{(n-1)}}\rangle
\langle\underline{p^{(n-1)}}|\,\varphi_{\pi n}(0)\rangle\nonumber\\
&  \times2\pi\delta\left(  k_{\pi2}^{1}-\sum\big(p_{j}^{(2)}\big)^{1}%
+\sum\big(p_{j}^{(1)}\big)^{1}\right)  \dots2\pi\delta\left(  k_{\pi n}%
^{1}+\sum\big(p_{j}^{(n-1)}\big)^{1}\right) \label{Xin}\\
&  \times\frac{-i}{\sum_{i=2}^{n}k_{\pi i}^{0}+\sum\omega_{j}^{(1)}-i\epsilon
}\frac{-i}{\sum_{i=3}^{n}k_{\pi i}^{0}+\sum\omega_{j}^{(2)}-i\epsilon}%
\dots\frac{-i}{k_{\pi n}^{0}+\sum\omega_{j}^{(n-1)}-i\epsilon}\nonumber
\end{align}
with $\int_{\underline{p^{(j)}}}=\int_{p_{1}^{(j)}}\dots\int_{p_{n_{j}}^{(j)}%
},~\int_{p}=\int\frac{d^{2}p}{2\pi2\omega},~\omega=\sqrt{m^{2}+\left(
p^{1}\right)  ^{2}},~\underline{n}!=\prod\limits_{j}\underline{n^{(j)}}%
!$,$~\underline{n^{(j)}}!=\prod\limits_{j}n_{k}^{(j)}!$ and $n_{k}^{(j)}=$ the
number of particles of type $k$ in the state $|\underline{p^{(j)}}\rangle$.
For explicit calculation it is convenient to take the limit $k_{i}%
^{1}\rightarrow0$, then the $\delta$-functions in (\ref{Xin}) simplify to%
\[
2\pi\delta\left(  \sum\big(p_{j}^{(1)}\big)^{1}\right)  \dots2\pi\delta\left(
\sum\big(p_{j}^{(n-1)}\big)^{1}\right)  \,.
\]

\subsubsection{S-matrix and form factors}

For integrable quantum field theories the n-particle S-matrix factorizes into
$n(n-1)/2$ two-particle ones
\[
S^{(n)}(\theta_{1},\dots,\theta_{n})=\prod_{i<j}S(\theta_{ij})\,,
\]
where the product on the right hand side has to be taken in a specific order
(see e.g.~\cite{KT}). The numbers $\theta_{ij}$ are the rapidity differences
$\theta_{ij}=\theta_{i}-\theta_{j}$, which are related to the momenta of the
particles by $p=m\left(  \cosh\theta,\sinh\theta\right)  $. To simplify the
calculations we will consider only theories with diagonal scattering and only
one type of particles. The generalization to more types of particles is
straightforward and will be used for the $Z_{3}$-model.

The form factors of a local bosonic field $\varphi(x)$ are the matrix elements%
\begin{equation}
F^{\varphi}(\underline{\theta})=\langle\,0\,|\,\varphi(0)\,|\,\theta_{1}%
,\dots,\theta_{n}\,\rangle\,.\label{F0}%
\end{equation}
They satisfy the form factor equations (i) -- (v) (see e.g. \cite{BFK}). As a
generalization we write
\[
F^{\varphi}(\underline{\theta}^{\prime};\underline{\theta})=\langle
\,\theta_{n^{\prime}}^{\prime},\dots,\theta_{1}^{\prime}\,|\,\varphi
(0)\,|\,\theta_{1},\dots,\theta_{n}\,\rangle
\]
which is related to (\ref{F0}) by crossing. In particular (see ref{a2})%
\begin{align}
&  F^{\varphi}(\theta_{1};\theta_{2},\theta_{3})=F^{\varphi}(\theta_{1}%
,\theta_{2}-i\pi_{-},\theta_{3}-i\pi_{+})+\delta_{\theta_{12}}+\delta
_{\theta_{13}}\label{F12}\\
&  F^{\varphi}(\theta_{2},\theta_{3};\theta_{4})=F^{\varphi}(\theta_{3}%
+i\pi_{+},\theta_{2}+i\pi_{-},\theta_{4})+\delta_{\theta_{24}}+\delta
_{\theta_{34}}\label{F21}%
\end{align}
with $i\pi_{\pm}=i\pi\pm i\epsilon$ and $\delta_{\theta_{12}}=4\pi
\delta(\theta_{1}-\theta_{2})$. The form factors $F^{\varphi}(\underline
{\theta})$ are meromorphic functions whereas the $F^{\varphi}(\underline
{\theta}^{\prime};\underline{\theta})$ are distributions.

\subsubsection{Examples}

\paragraph{The 2-point function}

Let $\varphi(x)$ be a scalar chargeless bosonic field with the normalization
$\langle0|\varphi(x)|p\rangle=1$. The 2-point Wightman function in the
1-particle intermediate state approximation is
\[
w^{1}(x_{1}-x_{2})=\int\frac{dp}{2\pi2\omega}\langle\,0\,|\,\varphi
(x_{1})\,|\,p\,\rangle\langle\,p\,|\,\varphi(x_{2})\,|\,0\rangle=i\Delta
_{+}\left(  x_{1}-x_{2}\right)
\]
and the 2-point Green's function in this approximation is%
\[
\tau^{1}(x_{1}-x_{2})=\Theta(x_{1}^{0}-x_{2}^{0})w^{1}(x_{1}-x_{2}%
)+\Theta(x_{2}^{0}-x_{1}^{0})w^{1}(x_{2}-x_{1})=i\Delta_{F}\left(  x_{1}%
-x_{2}\right)
\]
or in momentum space%
\begin{equation*}
\tilde{\tau}^{1}(k_{1},k_{2}) =\left(  2\pi\right)  ^{2}\delta
^{(2)}\left(  k_{1}+k_{2}\right)  \,\tilde{\Xi}(\underline{k})\,,~~~
\tilde{\Xi}(\underline{k}) =\frac{i}{k_{1}^{2}-m^{2}+i\epsilon}\,.
\end{equation*}

\paragraph{The 3-point function}

\label{s3p}

We consider $\tilde{\Xi}_{\underline{\varphi}}(\underline{k})$ for $n=3$. For
the detailed calculations see \ref{aZ2}. Let $\underline{\varphi}%
=(\varphi,\varphi,\epsilon)$ with $\langle0|\epsilon|\theta_{1},\theta
_{2}\rangle=F^{\epsilon}\left(  \theta_{1},\theta_{2}\right)  $. We calculate
$\tilde{\Xi}_{\varphi\varphi\epsilon}(\underline{k})$ in the limit $k_{i}%
^{1}\rightarrow0$. For the various permutations in (\ref{Xi}) we
obtain:\newline a) For the permutation $\pi=123$ we use the intermediate
states approximation \newline$\langle0|\varphi|\theta_{1}\rangle\langle
\theta_{1}|\varphi|\theta_{2},\theta_{3}\rangle\langle\theta_{3},\theta
_{2}|\epsilon|0\rangle$ then
\begin{align}
\tilde{\Xi}_{\varphi\varphi\epsilon}^{12}(k_{1},k_{2},k_{3})  &  =-\frac
{1}{64\pi m^{4}}\frac{m}{-k_{1}^{0}+m-i\epsilon}\int d\theta\frac{m}{\omega
}\frac{2m}{k_{3}^{0}+2\omega-i\epsilon}\label{Xi12}\\
&  \times F^{\varphi}(i\pi,\theta,-\theta)F^{\epsilon}(-\theta+i\pi
,\theta+i\pi)\nonumber
\end{align}
b) For the permutation $\pi=321$ we use the intermediate states approximation
\newline$\langle0|\epsilon|\theta_{1},\theta_{2}\rangle\langle\theta
_{2},\theta_{1}|\varphi|\theta_{3}\rangle\langle\theta_{3}|\varphi|0\rangle$
then%
\begin{align}
\tilde{\Xi}_{\epsilon\varphi\varphi}^{21}(k_{3},k_{2},k_{1})  &  =-\frac
{1}{64\pi m^{4}}\frac{m}{k_{1}^{0}+m-i\epsilon}\int d\theta\frac{m}{\omega
}\frac{2m}{-k_{3}^{0}+2\omega-i\epsilon}\label{Xi21}\\
&  \times F^{\epsilon}(\theta,-\theta)F^{\varphi}(-\theta+i\pi,\theta
+i\pi,0)\nonumber
\end{align}
c) For the permutation $\pi=132$ we use three intermediate states
approximations\newline i) $\left\langle 0|\varphi|\theta_{1}\rangle
\langle\theta_{1}|\epsilon|\theta_{2}\rangle\langle\theta_{2}|\varphi
|0\right\rangle :$%
\begin{equation}
\tilde{\Xi}_{\varphi\epsilon\varphi}^{11}(k_{1},k_{3},k_{2})=\frac{-1}{4m^{4}%
}\frac{m}{-k_{1}^{0}+m-i\epsilon}\frac{m}{k_{2}^{0}+m-i\epsilon}F^{\epsilon
}\left(  i\pi,0\right) \label{Xi11}%
\end{equation}
ii) $\langle0|\varphi|\theta_{1}\rangle\langle\theta_{1}|\epsilon|\theta
_{2},\theta_{3},\theta_{4}\rangle\langle\theta_{4},\theta_{3},\theta
_{2}|\varphi|0\rangle:$%
\begin{align}
\tilde{\Xi}_{\varphi\epsilon\varphi}^{13}(k_{1},k_{3},k_{2})  &  =-\frac
{1}{64\pi m^{4}}\frac{m}{-k_{1}^{0}+m-i\epsilon}\int d\theta\frac{m}{\omega
}\frac{2m}{k_{2}^{0}+2\omega+m-i\epsilon}\label{Xi13}\\
&  \times F^{\epsilon}(\theta,-\theta)F^{\varphi}(-\theta+i\pi,\theta
+i\pi,i\pi)\nonumber
\end{align}
iii) $\langle\,0\,|\,\varphi\,|\,\theta_{1},\theta_{2},\theta_{3}%
\rangle\,\langle\theta_{3},\theta_{2},\theta_{1}|\,\epsilon\,|\,\theta
_{4}\rangle\,\langle\theta_{4}\,|\,\varphi\,|\,0\,\rangle:$%
\begin{align*}
\tilde{\Xi}_{\varphi\epsilon\varphi}^{31}(k_{1},k_{3},k_{2})  &  =-\frac
{1}{64\pi m^{4}}\frac{m}{k_{2}^{0}+m-i\epsilon}\int d\theta\frac{m}{\omega
}\frac{2m}{-k_{1}^{0}+2\omega+m-i\epsilon}\\
&  \times F^{\varphi}(0,\theta,-\theta)F^{\epsilon}(-\theta+i\pi,\theta+i\pi)
\end{align*}
Finally we obtain
\begin{multline}
\tilde{\Xi}_{\varphi\varphi\epsilon}(k_{1},k_{2},k_{3})=\tilde{\Xi}%
_{\varphi\varphi\epsilon}^{12}(k_{1},k_{2},k_{3})+\tilde{\Xi}_{\epsilon
\varphi\varphi}^{21}(k_{3},k_{1},k_{2})\label{Xi3}\\
+\tilde{\Xi}_{\varphi\epsilon\varphi}^{11}(k_{1},k_{3},k_{2})+\tilde{\Xi
}_{\varphi\epsilon\varphi}^{13}(k_{1},k_{3},k_{2})+\tilde{\Xi}_{\varphi
\epsilon\varphi}^{31}(k_{1},k_{3},k_{2})+\left(  k_{1}\leftrightarrow
k_{2}\right)  .
\end{multline}

\paragraph{The 4-point function}

We consider $\tilde{\Xi}_{\underline{\varphi}}(\underline{k})$ for $n=4$. For
the detailed calculations see again \ref{a1}. Let $\underline{\varphi
}=(\varphi,\varphi,\varphi,\varphi)$. We use the intermediate states
approximation \newline$\langle0|\varphi|\theta_{1}\rangle\langle\theta
_{1}|\varphi|\theta_{2},\theta_{3}\rangle\langle\theta_{3},\theta_{2}%
|\varphi|\theta_{4}\rangle\langle\theta_{4}|\varphi|0\rangle$ then the
connected part yields for $k_{i}^{1}=0$
\begin{align}
\tilde{\Xi}_{\underline{\varphi}}(\underline{k})  &  =-\frac{1}{32}\frac
{i}{m^{6}\pi}\sum_{perm(k)}\frac{m}{-k_{1}^{0}+m-i\epsilon}\frac{m}{k_{4}%
^{0}+m-i\epsilon}g\left(  -\left(  k_{3}^{0}+k_{4}^{0}\right)  /(2m)+i\epsilon
\right) \label{Sigma4}\\
g(x)  &  =\frac{-1}{4}\int d\theta\frac{1}{\cosh\theta}\frac{1}{\cosh\theta
-x}I_{\underline{\varphi}}(0,\theta,-\theta,0)\nonumber
\end{align}
with%
\[
I_{\underline{\varphi}}(\theta_{1},\theta_{2},\theta_{3},\theta_{4}%
)=F^{\varphi}(\theta_{1};\theta_{2},\theta_{3})F^{\varphi}(\theta_{2}%
,\theta_{3};\theta_{4}))=I_{1}(\underline{\theta})+I_{2}(\underline{\theta})
\]
and%
\begin{align*}
I_{1}(\underline{\theta})  &  =\tfrac{1}{2}F^{\varphi}(\theta_{1},\theta
_{2}-i\pi_{+},\theta_{3}-i\pi_{-})F^{\varphi}(\theta_{3}+i\pi_{+},\theta
_{2}+i\pi_{-},\theta_{4})\\
&  +\tfrac{1}{2}F^{\varphi}(\theta_{1},\theta_{2}-i\pi_{-},\theta_{3}-i\pi
_{+})F^{\varphi}(\theta_{3}+i\pi_{-},\theta_{2}+i\pi_{+},\theta_{4})\\
I_{2}(\underline{\theta})  &  =\tfrac{1}{2}\left(  \delta_{\theta_{12}}\left(
1+S(\theta_{23})\right)  +\delta_{\theta_{13}}\left(  1+S(\theta_{23})\right)
\right)  F^{\varphi}(\theta_{3}+i\pi_{+},\theta_{2}+i\pi_{-},\theta_{4})\\
&  +\tfrac{1}{2}F^{\varphi}(\theta_{1},\theta_{2}-i\pi_{-},\theta_{3}-i\pi
_{+})\left(  \delta_{\theta_{24}}\left(  1+S(\theta_{32})\right)
+\delta_{\theta_{34}}\left(  1+S(\theta_{32})\right)  \right)
\end{align*}
see (\ref{I}).

\section{Models}

\subsection{The scaling $Z_{2}$ Ising model}

In the scaling limit this model may be described by an interacting Bose field
$\sigma_{n}^{z}=Cm^{1/8}\sigma(x)$, where $C$ is a numerical constant and
$m=h-J$. The excitations are non-interacting Majorana fermions with the
2-particle S-matrix $S(\theta)=-1$. The field $\epsilon(x)$ is defined by
$\sigma^{x}=(m/J)^{1/2}\epsilon(x)\sim\bar{\psi}\psi(x)$, where $\psi$ is a
free Majorana spinor field. The $n$-particle form factors for the order
parameter $\sigma(x)$ were proposed in \cite{BKW,K2} as%
\begin{equation}
F^{Z(2)}(\underline{\theta})=\langle\,0|\,\sigma(0)\,|\,\theta_{1}%
,\dots,\theta_{n}\,\rangle=\left(  2i\right)  ^{\frac{n-1}{2}}\prod_{i<j}%
\tanh\tfrac{1}{2}\theta_{ij}.\label{Fn}%
\end{equation}

\subsubsection{The 3-point function}

\begin{figure}
[th]
\begin{center}
\includegraphics[
natheight=2.483000in,
natwidth=3.724100in,
height=2.483in,
width=3.7241in
]%
{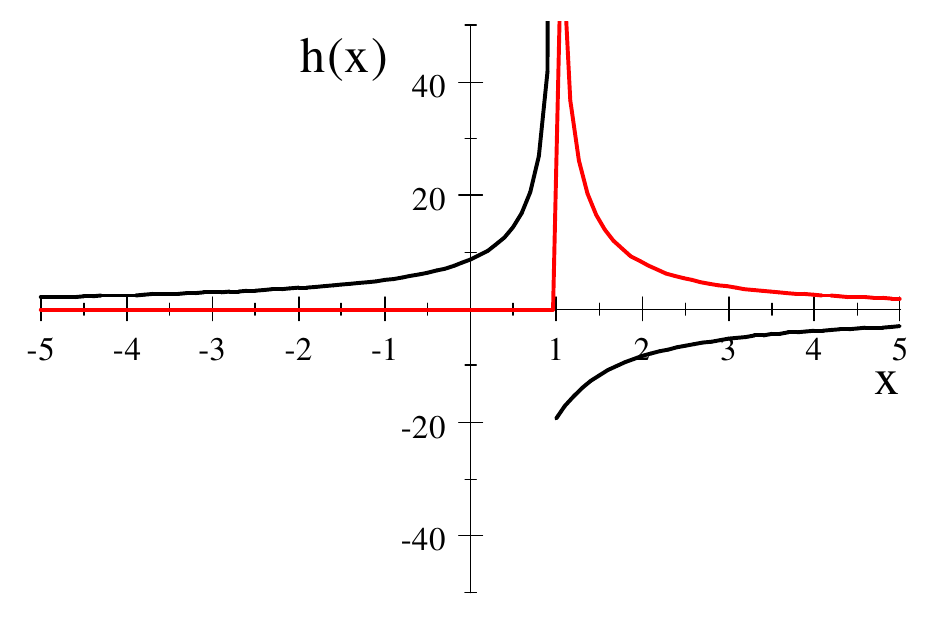}%
\caption{Plot of $\operatorname{Re}h_{+}^{Z2}(x)$ (black) and
$\operatorname{Im}h_{+}^{Z2}(x)$ (red).}%
\label{z2h+}%
\end{center}
\end{figure}

We investigate the Fourier transform of the Green's function
\[
\tau_{\varphi\varphi\epsilon}(\underline{x})=\langle0|T\varphi(x_{1}%
)\varphi(x_{2})\epsilon(x_{3})|0\rangle
\]
where $\varphi(x)$ is the order parameter $\sigma(x)$ and $\epsilon(x)\sim
\bar{\psi}\psi(x)$. For a free Majorana spinor field $\psi(x)$ we have (up to
a constant)
\[
\langle0|\epsilon(0)|\theta_{1},\theta_{2}\rangle=\sinh\tfrac{1}{2}\theta
_{12}\,.
\]
We apply the general results (\ref{Xi12}) -- (\ref{Xi3}) and obtain (for
details see \ref{aZ2})%
\begin{align*}
\tilde{\Xi}_{\varphi\varphi\epsilon}^{12}(k_{1},k_{2},k_{3})  &  =-\frac
{i}{32\pi m^{4}}\frac{m}{-k_{1}^{0}+m-i\epsilon}\,h_{+}^{Z2}(-k_{3}%
^{0}/(2m)+i\epsilon)\\
h_{+}^{Z2}(x)  &  =\frac{1}{2i}\int d\theta\frac{1}{\cosh\theta}\frac{1}%
{\cosh\theta-x}F^{\varphi}(i\pi,\theta,-\theta)F^{\epsilon}(-\theta
+i\pi,\theta+i\pi)
\end{align*}
Similarly we get%
\[
\tilde{\Xi}_{\epsilon\varphi\varphi}^{21}(k_{3},k_{2},k_{1})=-\frac{i}{32\pi
m^{4}}\frac{m}{k_{1}^{0}+m-i\epsilon}h_{+}^{Z2}(k_{3}^{0}/(2m)+i\epsilon)\,
\]
with (using (\ref{Fn}))
\begin{align*}
h_{+}^{Z2}(x)  &  =\int_{-\infty}^{\infty}\frac{1}{\cosh\theta}\frac{1}%
{\cosh\theta-x}\frac{\left(  \cosh\theta+1\right)  ^{2}}{\cosh\theta}d\theta\\
&  =-\frac{2}{x}-\frac{2}{x}\pi-\frac{1}{x^{2}}\pi-4\frac{\left(  1+x\right)
^{2}}{x^{2}\sqrt{x^{2}-1}}\operatorname{arctanh}\frac{1+x}{\sqrt{x^{2}-1}}\,.
\end{align*}
For the function $h_{+}^{Z2}(x)$ see Fig. \ref{z2h+}

\begin{figure}
[th]
\begin{center}
\includegraphics[
natheight=2.270400in,
natwidth=3.406000in,
height=2.2704in,
width=3.406in
]%
{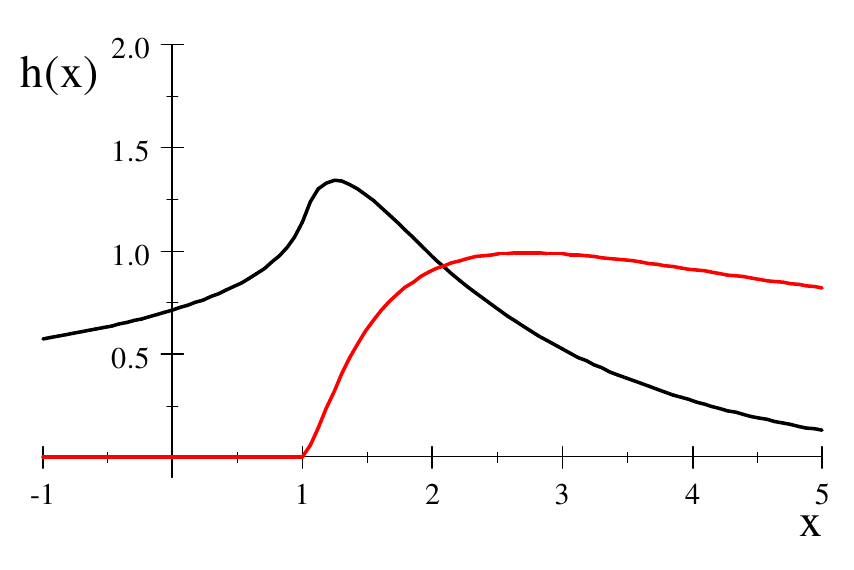}%
\caption{Plot of $\operatorname{Re}h_{-}^{Z2}(x)$ (black) and
$\operatorname{Im}h_{-}^{Z2}(x)$ (red).}%
\label{z2h-}%
\end{center}
\end{figure}

With (\ref{Xi13}) we obtain%
\begin{align*}
\tilde{\Xi}_{\varphi\epsilon\varphi}^{13}(k_{1},k_{3},k_{2})  &  =-\frac
{i}{32\pi m^{4}}\frac{m}{-k_{1}^{0}+m-i\epsilon}h_{-}^{Z2}(-k_{2}%
^{0}/(2m)-\tfrac{1}{2}+i\epsilon)\\
h_{-}^{Z2}(x)  &  =\frac{1}{2i}\int d\theta\frac{1}{\cosh\theta}\frac{1}%
{\cosh\theta-x}F^{\epsilon}(\theta,-\theta)F^{\varphi}(-\theta+i\pi
,\theta+i\pi,i\pi)
\end{align*}
and%
\[
\tilde{\Xi}_{\varphi\epsilon\varphi}^{31}(k_{1},k_{3},k_{2})=-\frac{i}{32\pi
m^{4}}\frac{m}{k_{2}^{0}+m-i\epsilon}h_{-}^{Z2}(k_{1}^{0}/(2m)-\tfrac{1}%
{2}+i\epsilon)
\]
with%
\begin{align*}
h_{-}^{Z2}(x)  &  =\int_{-\infty}^{\infty}d\theta\frac{\left(  \cosh
\theta-1\right)  ^{2}}{\cosh^{2}\theta}\frac{1}{\cosh\theta-x}\\
&  =-\frac{2}{x}+\frac{2}{x}\pi-\frac{1}{x^{2}}\pi-4\frac{\left(  x-1\right)
^{2}}{x^{2}\sqrt{x^{2}-1}}\operatorname{arctanh}\frac{1+x}{\sqrt{x^{2}-1}}\,.
\end{align*}
For the function $h_{-}^{Z2}(x)$ see Fig. \ref{z2h-}%

Finally%
\begin{align}
\tilde{\Xi}_{\varphi\varphi\epsilon}(k_{1},k_{2},k_{3})  &  =\tilde{\Xi
}_{\varphi\varphi\epsilon}^{12}(k_{1},k_{2},k_{3})+\tilde{\Xi}_{\epsilon
\varphi\varphi}^{21}(k_{3},k_{1},k_{2})+\tilde{\Xi}_{\varphi\epsilon\varphi
}^{11}(k_{1},k_{3},k_{2})\label{Xippe}\\
&  +\tilde{\Xi}_{\varphi\epsilon\varphi}^{13}(k_{1},k_{3},k_{2})+\tilde{\Xi
}_{\varphi\epsilon\varphi}^{31}(k_{1},k_{3},k_{2})+\left(  k_{1}%
\leftrightarrow k_{2}\right) \nonumber\\
&  =-\frac{i}{32\pi m^{4}}\frac{m}{-k_{1}^{0}+m-i\epsilon}h_{+}(-k_{3}%
^{0}/(2m)+i\epsilon)\nonumber\\
&  -\frac{i}{32\pi m^{4}}\frac{m}{k_{1}^{0}+m-i\epsilon}h_{+}(k_{3}%
^{0}/(2m)+i\epsilon)\nonumber\\
&  +\frac{-i}{4m^{4}}\frac{m}{-k_{1}^{0}+m-i\epsilon}\frac{m}{k_{2}%
^{0}+m-i\epsilon}\nonumber\\
&  -\frac{i}{32\pi m^{4}}\frac{m}{-k_{1}^{0}+m-i\epsilon}h_{-}(-k_{2}%
^{0}/(2m)-\tfrac{1}{2}+i\epsilon)\nonumber\\
&  -\frac{i}{32\pi m^{4}}\frac{m}{k_{2}^{0}+m-i\epsilon}h_{-}(k_{1}%
^{0}/(2m)-\tfrac{1}{2}+i\epsilon)+\left(  k_{1}\leftrightarrow k_{2}\right)
.\nonumber
\end{align}
This result can be applied to nonlinear susceptibility \cite{BKT}.

\subsubsection{The 4-point function}

\begin{figure}
[ptb]
\begin{center}
\includegraphics[
natheight=2.480500in,
natwidth=3.724100in,
height=2.4805in,
width=3.7241in
]%
{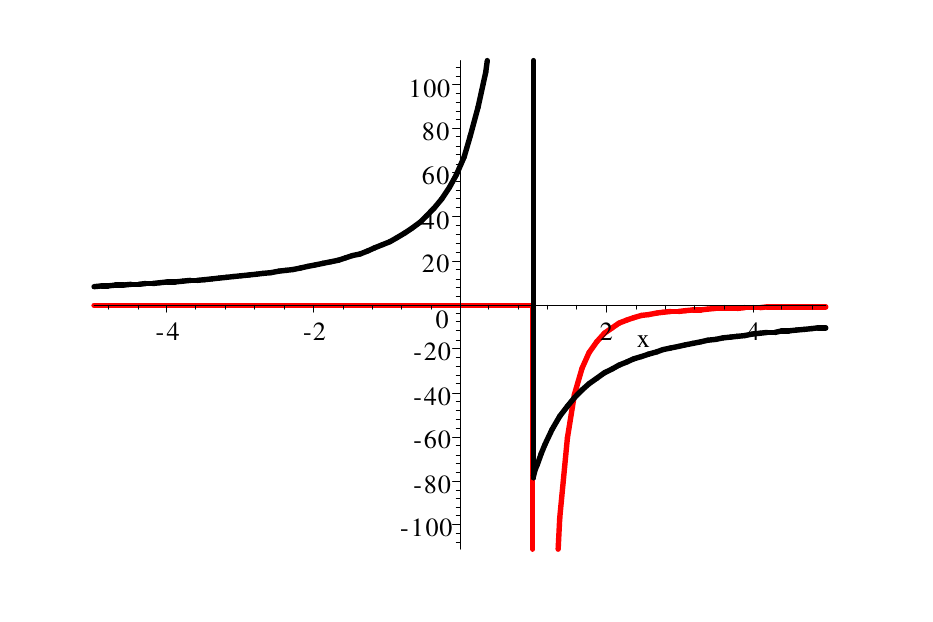}%
\caption{Plot of $\operatorname{Re}g^{Z2}(x)$ (black), $\operatorname{Im}%
g^{Z2}(x+i\epsilon)$ (red) for the scaling Ising model}%
\label{z2g}%
\end{center}
\end{figure}

We investigate the Fourier transform of the Green's function
\[
\tau_{\varphi\varphi\varphi\varphi}(\underline{x})=\langle0|T\varphi
(x_{1})\varphi(x_{2})\varphi(x_{3})\varphi(x_{4})|0\rangle
\]
for the order parameter $\varphi(x)=\sigma(x)$. From (\ref{Sigma4}) for
$k_{i}=(k_{i}^{0},0)$ in momentum space the contribution from $I_{2}$ in
(\ref{I}) vanishes, because $S(0)=-1$ and we get%
\begin{align}
\tilde{\Xi}_{\underline{\varphi}}(\underline{k})  &  =-\frac{i}{32\pi m^{6}%
}\sum_{perm(k)}\frac{m}{-k_{1}^{0}+m-i\epsilon}\frac{m}{k_{4}^{0}+m-i\epsilon
}g^{Z2}\left(  \frac{-1}{2m}\left(  k_{3}^{0}+k_{4}^{0}\right)  +i\epsilon
\right) \label{gZ2}\\
g^{Z2}(x)  &  =-\frac{1}{4}\int d\theta\frac{1}{\cosh\theta}\frac{1}%
{\cosh\theta-x}I_{\underline{\varphi}}^{Z_{2}}(0,\theta,-\theta,0)\nonumber
\end{align}
From (\ref{I}) and (\ref{Fn}) we obtain (see \ref{aZ2}) for the contribution
of $I_{1}$%
\begin{equation}
I_{\underline{\varphi}}^{Z_{2}}(0,\theta,-\theta,0)=2\tanh^{2}\theta\coth
^{4}\tfrac{1}{2}\left(  \theta-i\epsilon\right)  +\left(  \epsilon
\rightarrow-\epsilon\right) \label{IZ2}%
\end{equation}
and%
\begin{multline*}
g^{Z2}(x)=\frac{16}{1-x}-\frac{15\pi}{2x}-\frac{8}{x}-\frac{4\pi+2}{x^{2}%
}-\frac{\pi}{x^{3}}\\
-\frac{\left(  x+1\right)  ^{2}\sqrt{x^{2}-1}}{x^{3}\left(  x-1\right)  ^{2}%
}2\ln\left(  -x+\sqrt{x^{2}-1}\right)  \,.
\end{multline*}
For the function $g^{Z2}(x)$ see Fig. \ref{z2g}%

This result can be applied to Raman scattering \cite{BKT}.

\subsection{The sinh-Gordon model}

The classical field equation{\footnote{For details see \ref{asg}}} is%
\begin{equation}
\square\varphi(t,x)+\frac{\alpha}{\beta}\sinh\beta\varphi(t,x)=0.\label{eqm}%
\end{equation}

\textbf{The sinh-Gordon S-matrix }was derived in\cite{KT,Za2}\footnote{The
sinh-Gordon S-matrix is obtained from the sine-Gordon one by analytic
continuation of the coupling constant: $\beta\rightarrow i\beta$.}%
\[
S^{SG}(\theta)=\frac{\sinh\theta-i\sin\pi\mu}{\sinh\theta+i\sin\pi\mu}%
=-\exp\left(  -2\int_{0}^{\infty}\frac{dt}{t}\,\frac{\cosh\left(  \frac{1}%
{2}-\mu\right)  t}{\cosh\frac{1}{2}t}\sinh t\frac{\theta}{i\pi}\right)
\]
where $\mu$ is related to the coupling constant by%
\[
0<\mu=\frac{\beta^{2}}{8\pi+\beta^{2}}<1.
\]

\textbf{The sinh-Gordon minimal form factor} is \cite{KW,BK2}
\begin{equation}
F^{SG}(\theta)=\exp\int_{0}^{\infty}\frac{dt}{t\sinh t}\,\left(  \frac
{\cosh\left(  \frac{1}{2}-\mu\right)  t}{\cosh\frac{1}{2}t}-1\right)  \cosh
t\left(  1-\frac{\theta}{i\pi}\right)  \,.\nonumber
\end{equation}

\subsubsection{The 4-point function}

\begin{figure}
[ptb]
\begin{center}
\includegraphics[
natheight=1.928400in,
natwidth=2.896400in,
height=1.9284in,
width=2.8964in
]%
{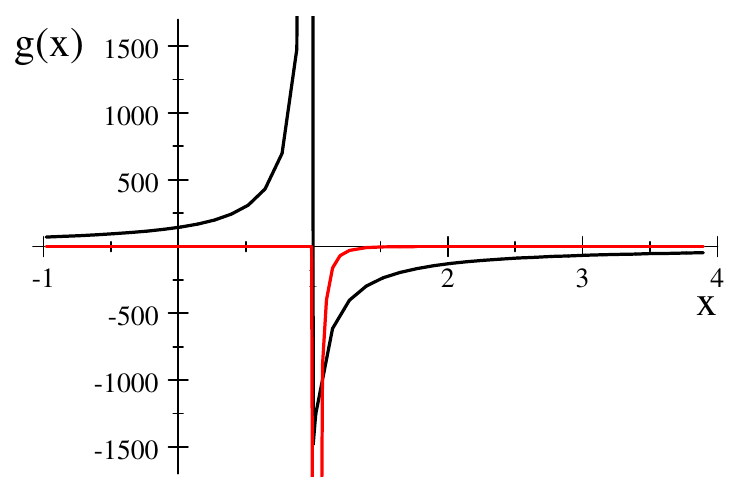}%
\caption{Plot of $\operatorname{Re}g_{1}^{SG}(x)$ (black) and
$\operatorname{Im}g_{1}^{SG}(x)$ (red) for $\mu=0.3$}%
\label{sgg3}%
\end{center}
\end{figure}

\begin{figure}
[ptb]
\begin{center}
\includegraphics[
natheight=1.992200in,
natwidth=2.991700in,
height=1.9922in,
width=2.9917in
]%
{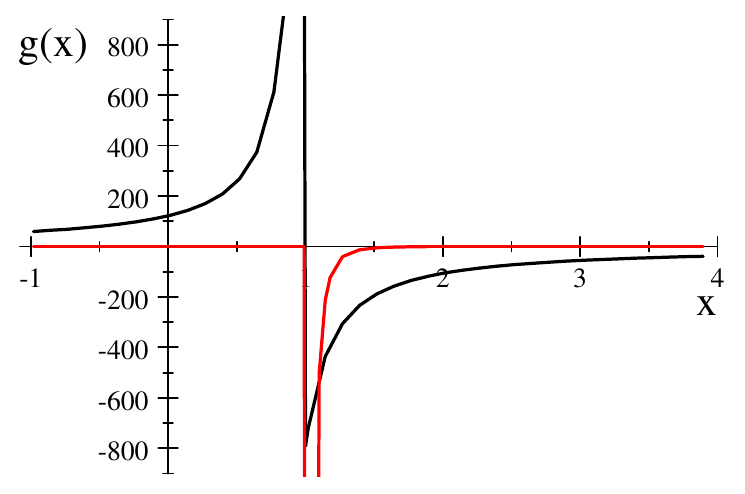}%
\caption{Plot of $\operatorname{Re}g_{1}^{SG}(x)$ (black) and
$\operatorname{Im}g_{1}^{SG}(x)$ (red) for $\mu=0.5$}%
\label{sgg5}%
\end{center}
\end{figure}

We consider the 4-point Green's function
\[
\tau_{\varphi\varphi\varphi\varphi}(\underline{x})=\langle\,0\,|\,T\varphi
(x_{1})\varphi(x_{2})\varphi(x_{3})\varphi(x_{4})|\,0\,\rangle
\]
and calculate the function $\tilde{\Xi}_{\underline{\varphi}}(\underline{k}) $
given by (\ref{Sigma4}) (for details see \ref{asg})%
\[
\tilde{\Xi}_{\underline{\varphi}}^{SG}(\underline{k})=\frac{-i}{32\pi m^{6}%
}\sum_{perm(k)}\frac{m}{-k_{1}^{0}+m-i\epsilon}\frac{m}{k_{4}^{0}+m-i\epsilon
}g^{SG}\left(  \frac{-1}{2m}\left(  k_{3}^{0}+k_{4}^{0}\right)  \right)
\]
with $g^{SG}(x)=g_{1}^{SG}(x)+g_{2}^{SG}(x)$ and%
\[
g_{i}^{SG}(x)=-\frac{1}{4}\int\frac{1}{\cosh\theta}\frac{1}{\cosh\theta
-x}I_{\underline{\varphi}i}^{SG}(0,\theta,-\theta,0)d\theta
\]
From (\ref{I}) and (\ref{Fsg3}) we obtain using $I_{\underline{\varphi}%
}^{Z_{2}}$ as defined in (\ref{IZ2})%
\begin{align*}
&  I_{\underline{\varphi}1}^{SG}(0,\theta,-\theta,0)\\
&  =\tfrac{1}{2}Z^{\varphi}F^{SG}(0,\theta-i\pi_{+},-\theta-i\pi_{-}%
)F^{SG}(-\theta+i\pi_{+},\theta+i\pi_{-},0)+\left(  \epsilon\rightarrow
-\epsilon\right) \\
&  =f^{SG}(\theta)I_{\underline{\varphi}}^{Z_{2}}(0,\theta,-\theta,0)
\end{align*}
where\footnote{As usual, in the context of the sine-Gordon model, the
normalization of the field is given by $\langle\,0\,|\,\varphi
(0)\,|\,p\,\rangle=\sqrt{Z^{\varphi}}$ (see \ref{asg}).}%
\[
f^{SG}(\theta)=-\frac{\left(  Z^{\varphi}\right)  ^{2}\sin^{2}\pi\mu}%
{F^{2}\left(  i\pi\right)  \left(  2i\right)  ^{2}}\left(  F_{0}(\theta
+i\pi)\right)  ^{4}F_{0}(2\theta)F_{0}(-2\theta)
\]
and $F_{0}(\theta)=F^{SG}(\theta)/\left(  -i\sinh\tfrac{1}{2}\theta\right)  $.
Therefore as in (\ref{gZ2}) we obtain%
\[
g_{1}^{SG}(x)=-\int_{-\infty}^{\infty}f^{SG}(\theta)\left(  \frac{\coth
^{4}\tfrac{1}{2}\theta\,\tanh^{2}\theta}{\cosh\theta}\frac{1}{\cosh\theta
-x}-\frac{16}{\theta^{2}}\frac{1}{1-x}\right)  d\theta.
\]
The functions $g_{1}^{SG}(x)$ for $\mu=0.3$ and $\mu=0.5$ are plotted in Fig.
\ref{sgg3} and \ref{sgg5}.%
The function $g_{2}^{SG}\left(  x\right)  $ given by $I_{\underline{\varphi}%
2}(\underline{\theta})$ as defined in (\ref{I}) follows from (\ref{g2})%
\begin{equation}
g_{2}^{SG}\left(  x\right)  =-32\frac{\pi\sqrt{Z^{\varphi}}}{\sin\pi\mu}%
\frac{1}{1-x}.\label{gsg2}%
\end{equation}

\subsection{The {$Z_{3}$}-model}

The model we consider is the $Z_{N}$-symmetric CFT perturbed by the thermal
operator
\begin{equation}
S=S[Z_{N}]+\lambda\int d\tau dx\epsilon(\tau,x)\label{Z3}%
\end{equation}
for a particular value $N=3$. Such model appears as the continuum limit of the
lattice model describing the integrable anti-ferromagnetic chain of spins
$S=N/2$ in an applied magnetic field \cite{tsvNP}
\begin{equation}
H=\sum_{n}\Big[JP_{N}(\mathbf{S}_{n}\mathbf{S}_{n+1})+HS_{n}^{z}%
\Big],\label{Hratch}%
\end{equation}
where $P_{N}(x)$ is the polynomial of the $N$-th degree \cite{Takh}%
,\cite{Bab}. The continuum limit of this model at $H=0$ is the $SU_{N}(2)$
Wess-Zumino-Novikov-Witten (WZNW) model perturbed by the irrelevant operator
\[
H=W[SU_{N}(2)]+\eta\int dxJ^{a}\bar{J}^{b}\Phi_{adj}^{ab},
\]
where $\Phi_{adj}$ is the primary field in the adjoint representation and
$J,\bar{J}$ are the holomorphic and antiholomorphic currents of the
$su_{N}(2)$ Kac-Moody algebra. Whence the magnetic field is applied along the
$z$-axis the $z$-components of the currents acquire finite expectation
values
\[
\langle\,J^{z}\,\rangle=\langle\,\bar{J}^{z}\,\rangle=\frac{1}{2}\chi H,
\]
where $\chi\sim1/J$ is the uniform magnetic susceptibility, and the irrelevant
operator becomes relevant \cite{tsv2000}
\[
J^{a}\bar{J}^{b}\Phi_{adj}^{ab}\rightarrow\frac{1}{4}(\chi H)^{2}\Phi
_{adj}^{zz}.
\]
The conformal embedding $SU_{N}(2)=U(1)\times Z_{N}$ establishes the
equivalence between the diagonal component of the adjoint primary field and
the thermal operator $\epsilon$ and hence the equivalence between the massive
sector of model (\ref{Hratch}) and model (\ref{Z3}).

The $Z_{3}$ CFT and the exact solution of the massive theory (\ref{Z3}) for
$N=3$ suggest that in the disordered phase there are 2 types of particles $1$
and $2$ and two corresponding fields (order parameters $\sigma_{1},~\sigma
_{2}=\sigma_{1}^{\ast}$) with%
\[
\langle\,0\,|\,\sigma_{1}(x)\,|\,p\rangle_{1}=1\,,~~\langle\,0\,|\,\sigma
_{2}(x)\,|\,p\rangle_{2}=1\,.
\]
where the indices correspond to the emission of particle 1 and 2 (the latter
is a bound state of two 1-particles and simultaneously the anti-particle of
particle 1).

\textbf{The two-particle S-matrix} for the $Z_{3}$-Potts model perturbed by
the thermal operator has been proposed by K\"{o}berle and Swieca \cite{KS}. It
coincides with the one derived from the Bethe ansatz solution of model
(\ref{Hratch}) \cite{tsvNP}. The scattering matrix of two particles of type $1
$ is
\[
S^{Z_{3}}(\theta)=\frac{\sinh\frac{1}{2}(\theta+\frac{2}{3}i\pi)}{\sinh
\frac{1}{2}(\theta-\frac{2}{3}i\pi)}\,.
\]
This S-matrix is consistent with the picture that the bound state of two
particles of type $1$ is the particle $2$ which is the anti-particle of $1$.

\textbf{The form factors of the }$Z_{N}$\textbf{-model} (\ref{Z3}) have been
proposed in \cite{K3,KiSm,BFK}. The minimal solution of the Watson's and the
crossing equations
\[
F(\theta)=F(-\theta)S(\theta)\,,~~F(i\pi-\theta)=F(i\pi+\theta)
\]
for the $Z(3)$ model is%
\begin{align*}
F^{Z3}(i\pi x)  &  =\sin\tfrac{1}{2}\pi x\,\exp\int_{0}^{\infty}\left(
\frac{\sinh\frac{1}{3}t}{t\sinh^{2}t}\left(  1-\cosh t\left(  1-x\right)
\right)  \right)  dt\\
&  =\sin\tfrac{1}{2}\pi x\,\frac{G\left(  \frac{1}{3}+\frac{1}{2}x\right)
G\left(  \frac{4}{3}-\frac{1}{2}x\right)  }{G\left(  \frac{2}{3}+\frac{1}%
{2}x\right)  G\left(  \frac{5}{3}-\frac{1}{2}x\right)  }%
\end{align*}
where $G(x)$ is the Barnes G-function \cite{Wo} with the defining relation
\[
G\left(  x+1\right)  =G\left(  x\right)  \Gamma\left(  x\right)  \,.
\]
The form factor of the order parameter field $\sigma_{1}$ and two particles of
type $2$ is
\begin{equation}
F_{22}^{\sigma_{1}}\left(  \underline{\theta}\right)  =\langle\,0\,|\,\sigma
_{1}(0)\,|\,p_{1},p_{2}\,\rangle_{22}=c_{2}\frac{F(\theta_{12})}{\sinh\frac
{1}{2}(\theta_{12}-\frac{2}{3}i\pi)\sinh\frac{1}{2}\left(  \theta_{12}%
+\frac{2}{3}i\pi\right)  }\label{F22}%
\end{equation}
and for the 3 particles of type $1,1$ and $2$%
\begin{align}
F_{112}^{\sigma_{1}}\left(  \underline{\theta}\right)   &  =\langle
\,0\,|\,\sigma_{1}(0)\,|\,p_{1},p_{2},p_{3}\,\rangle_{112}\label{F112}\\
&  =c_{3}\frac{F(\theta_{12})\cosh\frac{1}{2}\theta_{12}}{\sinh\frac{1}%
{2}(\theta_{12}-\frac{2}{3}i\pi)\sinh\frac{1}{2}\left(  \theta_{12}+\frac
{2}{3}i\pi\right)  }\prod_{i=1}^{2}\frac{F_{12}^{\min}(\theta_{i3})}%
{\cosh\frac{1}{2}\theta_{i3}}\nonumber
\end{align}
where
\begin{align*}
F_{(12)}^{\min}(i\pi x)  &  =c\exp\int_{0}^{\infty}\frac{dt}{t\sinh^{2}t}%
\sinh\frac{2}{3}t\,\left(  1-\cosh t\left(  1-x\right)  \right) \\
&  =\frac{G\left(  \frac{1}{6}+\frac{1}{2}x\right)  G\left(  \frac{7}{6}%
-\frac{1}{2}x\right)  }{G\left(  \frac{5}{6}+\frac{1}{2}x\right)  G\left(
\frac{11}{6}-\frac{1}{2}x\right)  }%
\end{align*}
is the minimal form factor of the particles $1$ and $2$.

\subsubsection{The 3-point function}

We consider the Green's function $\tau_{\sigma_{1}\sigma_{1}\sigma_{1}%
}(\underline{x})=\langle\,0\,|\,T\sigma_{1}(x_{1})\sigma_{1}(x_{2})\sigma
_{1}(x_{3})\,|\,0\,\rangle$, this 3-point function was also investigated in
\cite{CDGJM}. As in (\ref{Xi11}) we have the simple contribution%
\begin{align*}
\tilde{\Xi}_{\sigma_{1}\sigma_{1}\sigma_{1}}^{11}(k_{1},k_{2},k_{3})  &
=\int_{p_{1}}\int_{p_{2}}\left\langle 0|\sigma_{1}(0)|p_{1}\rangle\langle
p_{1}|\sigma_{1}(0)|\bar{p}_{2}\rangle\langle\bar{p}_{2}|\,\sigma
_{1}(0)|0\right\rangle \\
&  \times2\pi\delta\left(  p_{1}\right)  2\pi\delta\left(  \bar{p}_{2}\right)
\frac{-i}{k_{2}^{0}+k_{3}^{0}+\omega_{1}-i\epsilon}\frac{-i}{k_{3}^{0}%
+\omega_{2}-i\epsilon}\\
&  =\frac{-1}{4m^{4}}\frac{m}{-k_{1}^{0}+m-i\epsilon}\frac{m}{k_{3}%
^{0}+m-i\epsilon}F_{22}^{\sigma_{1}}(i\pi,0)
\end{align*}
and as in (\ref{Xi12}) we calculate for the intermediate states \newline%
$\left\langle 0|\sigma_{1}(0)|p_{1}\rangle\langle p_{1}|\sigma_{1}%
(0)|p_{2},p_{3}\rangle\langle p_{3},p_{2}|\,\sigma_{1}(0)|0\right\rangle $
(for details see \ref{aZ3})
\begin{align*}
\tilde{\Xi}_{\sigma_{1}\sigma_{1}\sigma_{1}}^{12}(k_{1},k_{2},k_{3})  &
=-\frac{1}{64\pi m^{4}}\frac{m}{-k_{1}^{0}+m-i\epsilon}\int d\theta\frac
{m}{\omega}\frac{2m}{k_{3}^{0}+2\omega-i\epsilon}I_{\sigma_{1}\sigma_{1}%
\sigma_{1}}^{12}(\theta)\\
I_{\sigma_{1}\sigma_{1}\sigma_{1}}^{12}(\theta)  &  =F_{211}^{\sigma_{1}}%
(i\pi,\theta,-\theta)F_{22}^{\sigma_{1}}(-\theta+i\pi,\theta+i\pi)
\end{align*}
where we have used the crossing relation%
\[
\langle p_{1}|\sigma_{1}(0)|p_{2},p_{3}\rangle=F_{211}^{\sigma_{1}}(\theta
_{1}+i\pi,\theta_{2},\theta_{3})+\delta_{\theta_{1}\theta_{2}}+\delta
_{\theta_{1}\theta_{3}}S(\theta_{23}).
\]
The $\delta$-function terms do not contribute because $F_{22}^{\sigma_{1}%
}(0,0)=0$. Inserting the form factor functions we get (up to constant factors)%
\begin{gather*}
\tilde{\Xi}_{\sigma_{1}\sigma_{1}\sigma_{1}}^{12}(k_{1},k_{2},k_{3})=-\frac
{1}{64\pi m^{4}}\frac{m}{-k_{1}^{0}+m-i\epsilon}h^{Z3}\left(  -\frac{k_{3}%
^{0}}{2m}+i\epsilon\right) \\
h^{Z3}\left(  x\right)  =\int_{-\infty}^{\infty}d\theta\frac{1}{\cosh\theta
}\frac{1}{\cosh\theta-x}\frac{F(2\theta)F(-2\theta)F_{(12)}^{\min}(\theta
+i\pi)F_{(12)}^{\min}(-\theta+i\pi)}{\left(  \sinh(\theta-\frac{1}{3}%
i\pi)\,\sinh(\theta+\frac{1}{3}i\pi)\,\sinh\frac{1}{2}\theta\right)  ^{2}}.
\end{gather*}
For the intermediate states $\left\langle 0|\sigma_{1}(0)|\bar{p}_{1},\bar
{p}_{2}\rangle\langle\bar{p}_{2},\bar{p}_{1}|\sigma_{1}(0)|\bar{p}_{3}%
\rangle\langle\bar{p}_{3}|\,\sigma_{1}(0)|0\right\rangle $ we get%
\[
\tilde{\Xi}_{\sigma_{1}\sigma_{1}\sigma_{1}}^{\bar{2}\bar{1}}(k_{1}%
,k_{2},k_{3})=-\frac{1}{64\pi m^{4}}\frac{m}{k_{3}^{0}+m-i\epsilon}%
h^{Z3}\left(  \frac{k_{1}^{0}}{2m}+i\epsilon\right)
\]
and as in (\ref{Xi3})
\begin{align*}
\tilde{\Xi}_{\sigma_{1}\sigma_{1}\sigma_{1}}(k_{1},k_{2},k_{3})  &
=const.\sum_{perm(k)}\frac{m}{-k_{1}^{0}+m-i\epsilon}\frac{m}{k_{3}%
^{0}+m-i\epsilon}\\
& +const.^{\prime}\sum_{perm(k)}\frac{m}{-k_{1}^{0}+m-i\epsilon}%
h^{Z(3)}\left(  -\frac{k_{3}^{0}}{2m}+i\epsilon\right)  +\left(
k_{i}\rightarrow-k_{i}\right)  \,.
\end{align*}
\begin{figure}
[th]
\begin{center}
\includegraphics[
natheight=2.406500in,
natwidth=3.616900in,
height=2.4065in,
width=3.6169in
]%
{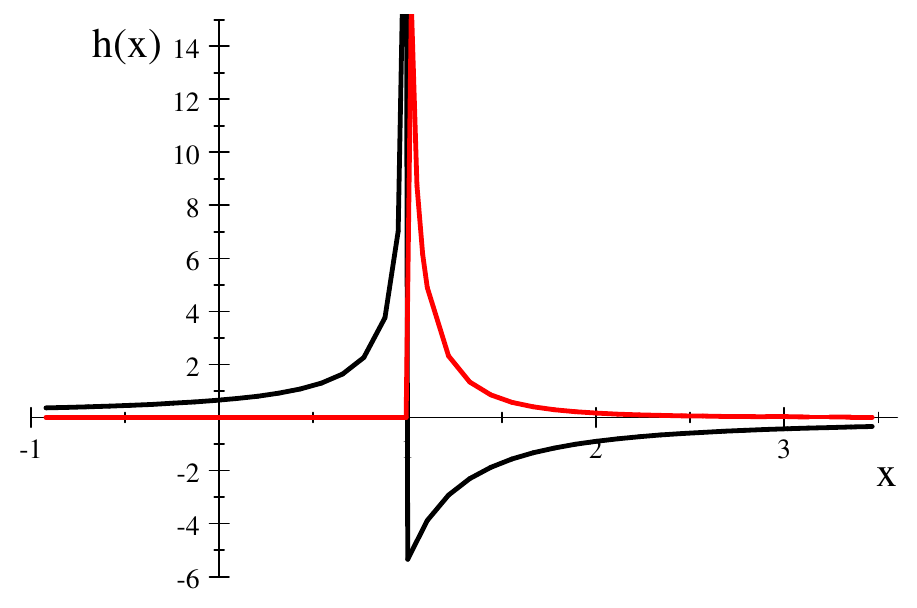}%
\caption{Plot of $\operatorname{Re}h^{Z3}(x)$ (black) and $\operatorname{Im}%
h^{Z3}(x)$ (red) for the $Z_{3}$ model}%
\label{hz3}%
\end{center}
\end{figure}
As expected, there is a threshold singularity at $k^{0}=2m$.

\subsubsection{The 4-point function}

\begin{figure}
[t]
\begin{center}
\includegraphics[
natheight=2.088300in,
natwidth=3.130400in,
height=2.0883in,
width=3.1304in
]%
{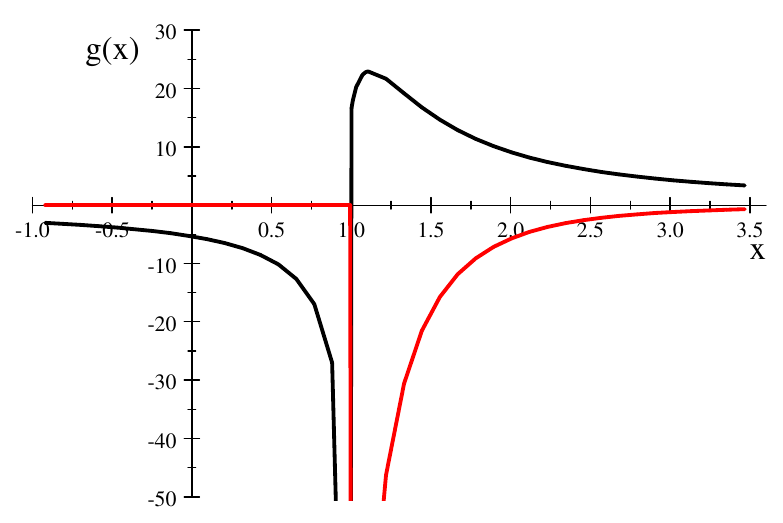}%
\caption{Plot of $\operatorname{Re}g_{I}^{Z3}(x)$ (black) and
$\operatorname{Im}g_{I}^{Z3}(x)$ (red) for the $Z_{3}$ model}%
\label{gz31}%
\end{center}
\end{figure}

\begin{figure}
[th]
\begin{center}
\includegraphics[
natheight=2.097700in,
natwidth=3.144000in,
height=2.0977in,
width=3.144in
]%
{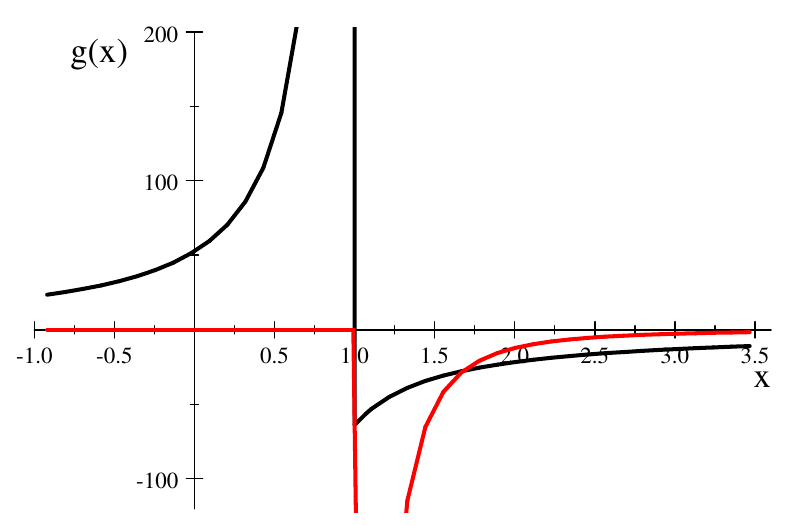}%
\caption{Plot of $\operatorname{Re}g_{II}^{Z3}(x)$ (black) and
$\operatorname{Im}g_{II}^{Z3}(x) $ (red) for the $Z_{3}$ model }%
\label{gz32}%
\end{center}
\end{figure}

Due to (\ref{I}) there are contributions to the four point Green's function
from $I_{1},I_{2}$ and $I_{3}$. The one from $I_{3}$ belongs to the
disconnected part and the one from $I_{2}$ is trivial as in (\ref{gsg2}) and
(\ref{g2}). We restrict here to the contribution from $I_{1}$. We consider the
Green's function $\tau_{\sigma_{1}\sigma_{2}\sigma_{1}\sigma_{2}}%
(\underline{x})=\langle\,0\,|\,T\sigma_{1}(x_{1})\sigma_{2}(x_{2})\sigma
_{1}(x_{3})\sigma_{2}(x_{4})\,|\,0\,\rangle$ and as in (\ref{Sigma4}) we
obtain (for details see \ref{aZ3})%
\begin{multline*}
\tilde{\Xi}_{\sigma_{1}\sigma_{2}\sigma_{3}\sigma_{4}}(\underline{k}%
)=\sum_{\pi\in S_{4}}\tilde{\Xi}_{\sigma_{\pi1}\sigma_{\pi2}\sigma_{\pi
3}\sigma_{\pi4}}^{121}(k_{\pi1},k_{\pi2},k_{\pi3},k_{\pi4})\,,~~\left(
\sigma_{3}=\sigma_{1},\sigma_{4}=\sigma_{2}\right) \\
=-\frac{1}{32}\frac{i}{m^{6}\pi}\sum_{\pi\in S_{4}}\frac{m}{-k_{\pi1}%
^{0}+m-i\epsilon}\frac{m}{k_{\pi4}^{0}+m-i\epsilon}g_{\pi}^{Z3}\left(
-\frac{k_{\pi3}^{0}+k_{\pi4}^{0}}{2m}+i\epsilon\right)  .
\end{multline*}
Obviously, if $\sigma_{3}=\sigma_{1}$ and $\sigma_{4}=\sigma_{2}$ there are
three functions $g_{\pi}^{Z3}(x)$%
\[
g_{\pi}^{Z3}(x)=\left\{
\begin{array}
[c]{lll}%
g_{I}^{Z3}(x) & \text{if} & \sigma_{\pi1}\sigma_{\pi2}\sigma_{\pi3}\sigma
_{\pi4}=\sigma_{1}\sigma_{2}\sigma_{1}\sigma_{2}~\text{or }\left(  \sigma
_{1}\leftrightarrow\sigma_{2}\right) \\
g_{II}^{Z3}(x) & \text{if} & \sigma_{\pi1}\sigma_{\pi2}\sigma_{\pi3}%
\sigma_{\pi4}=\sigma_{1}\sigma_{1}\sigma_{2}\sigma_{2}~\text{or }\left(
\sigma_{1}\leftrightarrow\sigma_{2}\right) \\
g_{III}^{Z3}(x) & \text{if} & \sigma_{\pi1}\sigma_{\pi2}\sigma_{\pi3}%
\sigma_{\pi4}=\sigma_{1}\sigma_{2}\sigma_{2}\sigma_{1}~\text{or }\left(
\sigma_{1}\leftrightarrow\sigma_{2}\right)  \,.
\end{array}
\right.
\]
It turns out that $g_{III}^{Z3}(x)=g_{I}^{Z3}(x)$. For plots of the functions
$g_{I}^{Z3}(x)$ and $g_{II}^{Z3}(x)$ see Figs. \ref{gz31} and \ref{gz32}.%

\section*{Conclusion}

In this paper we develop a technique to calculate multipoint Wightman or Green
functions in integrable quantum field theories in 1+1 dimension. We insert
intermediate states between the fields and use the crossing formula to write
the Wightman function in terms of form factors in a model independent way. We
expect good approximations for low number of particles in the intermediate
states. In the present article we demonstrate this technique explicitly for 3-
and 4-point functions of simple models with no backward scattering: the
scaling $Z_{2}$ Ising, the scaling $Z_{3}$ Potts and the sinh-Gordon model.
The results can be applied to physical phenomena, for example to Raman
scattering \cite{BKT}. In a forthcoming article we will generalize the
technique to models with backward scattering, as the $O(N)~\sigma$- and the
$O(N)$ Gross-Neveu model.

\subsection*{Acknowledgments}


We are grateful to G. Blumberg, J. Misewich and especially to N. P. Armitage
for advising us on the experimentally related matters, to S. Lukyanov who
pointed out for us paper \cite{BNNPSW} and to A. B. Zamolodchikov for fruitful
discussions. A.~M.~T. was supported by the U.S. Department of Energy (DOE),
Division of Materials Science, under Contract No. DE-AC02-98CH10886. H.~B. is
grateful to Simons Center and IIP in Natal for hospitality and support. H.~B.
also supported by Armenian grant 15T-1C308 and by ICTP OEA-AC-100 project.
M.~K. was supported by Fachbereich Physik, Freie Universit\"{a}t Berlin.

\appendix

\section{Crossing}

\label{a2}

The general crossing formula (31) in \cite{BK}%
\begin{align*}
&  F^{\mathcal{O}}{}_{I}^{J}(\underline{\theta}_{J}^{\prime};\underline
{\theta}_{I})\\
&  =\sigma_{\mathcal{O}J}\sum_{%
\genfrac{}{}{0pt}{1}{L\cup N=J}{K\cup M=I}%
}\dot{S}_{NL}^{J}(\underline{\theta}_{N}^{\prime},\underline{\theta}%
_{L}^{\prime})\,\mathbf{1}_{M}^{N}(\underline{\theta}_{N}^{\prime}%
,\underline{\theta}_{M})\,\mathbf{C}^{L\bar{L}}F_{\bar{L}K}^{\mathcal{O}%
}(\underline{\theta}_{\bar{L}}^{\prime}+i\pi_{-},\underline{\theta}_{K}%
)\,\dot{S}_{I}^{MK}(\underline{\theta}_{I})\\
&  =\sum_{%
\genfrac{}{}{0pt}{1}{L\cup N=J}{K\cup M=I}%
}\dot{S}_{LN}^{J}(\underline{\theta}_{L}^{\prime},\underline{\theta}%
_{N}^{\prime})\,F_{K\bar{L}}^{\mathcal{O}}(\underline{\theta}_{K}%
,\underline{\theta}_{\bar{L}}^{\prime}-i\pi_{-})\mathbf{C}^{\bar{L}%
L}\,\mathbf{1}_{M}^{N}(\underline{\theta}_{N}^{\prime},\underline{\theta}%
_{M})\,\dot{S}_{I}^{KM}(\underline{\theta}_{I})
\end{align*}%
\[%
\begin{array}
[c]{l}%
\unitlength.22mm\begin{picture}(60,70)(10,0) \put(30,35){\oval(60,30)[]} \put(10,0){\line(0,1){20}} \put(50,0){\line(0,1){20}} \put(10,50){\line(0,1){20}} \put(50,50){\line(0,1){20}} \put(30,35){\makebox(0,0)[cc]{$\cal O$}} \put(30,5){\makebox(0,0)[cc]{$I$}} \put(30,65){\makebox(0,0)[cc]{$J$}} \end{picture}
\end{array}
=\sum_{%
\genfrac{}{}{0pt}{1}{N\cup L=J}{M\cup K=I}%
}%
\begin{array}
[c]{l}%
\unitlength0.22mm\begin{picture}(140,135) \put(110,85){\oval(60,30)[]} \put(110,85){\makebox(0,0)[cc]{$\cal O$}} \put(81,70){\oval(52,50)[b]} \put(80,70){\oval(20,30)[b]} \put(130,40){\line(0,1){30}} \put(115,40){\line(0,1){30}} \put(70,70){\line(0,1){35}} \put(55,70){\line(0,1){35}} \put(110,10){\line(0,1){10}} \put(40,10){\line(0,1){10}} \put(110,125){\line(0,1){10}} \put(40,125){\line(0,1){10}} \put(75,10){\makebox(0,0)[cc]{$I$}} \put(75,135){\makebox(0,0)[cc]{$J$}} \put(122.50,55){\makebox(0,0)[cc]{$K$}} \put(95,61){\makebox(0,0)[cc]{$\bar L$}} \put(62,85){\makebox(0,0)[cc]{$L$}} \put(5,20){\framebox(135,20)[cc]{$\dot S$}} \put(5,105){\framebox(135,20)[cc]{$\dot S$}} \put(25,72.50){\oval(40,25)[]} \put(25,72.50){\makebox(0,0)[cc]{$\bf 1$}} \put(15,40){\line(0,1){20}} \put(35,40){\line(0,1){20}} \put(35,85){\line(0,1){20}} \put(15,85){\line(0,1){20}} \put(25,50){\makebox(0,0)[cc]{$M$}} \put(25,95){\makebox(0,0)[cc]{$N$}} \end{picture}
\end{array}
=\sum_{%
\genfrac{}{}{0pt}{1}{L\cup N=J}{K\cup M=I}%
}%
\begin{array}
[c]{l}%
\unitlength0.22mm\begin{picture}(135,135) \put(30,85){\oval(60,30)[]} \put(30,85){\makebox(0,0)[cc]{$\cal O$}} \put(59,70){\oval(52,50)[b]} \put(60,70){\oval(20,30)[b]} \put(10,40){\line(0,1){30}} \put(25,40){\line(0,1){30}} \put(70,70){\line(0,1){35}} \put(85,70){\line(0,1){35}} \put(30,10){\line(0,1){10}} \put(100,10){\line(0,1){10}} \put(30,125){\line(0,1){10}} \put(100,125){\line(0,1){10}} \put(65,10){\makebox(0,0)[cc]{$I$}} \put(65,135){\makebox(0,0)[cc]{$J$}} \put(17.50,55){\makebox(0,0)[cc]{$K$}} \put(45,61){\makebox(0,0)[cc]{$\bar L$}} \put(78,85){\makebox(0,0)[cc]{$L$}} \put(0,20){\framebox(135,20)[cc]{$\dot S$}} \put(0,105){\framebox(135,20)[cc]{$\dot S$}} \put(115,72.50){\oval(40,25)[]} \put(115,72.50){\makebox(0,0)[cc]{$\bf 1$}} \put(125,40){\line(0,1){20}} \put(105,40){\line(0,1){20}} \put(105,85){\line(0,1){20}} \put(125,85){\line(0,1){20}} \put(115,50){\makebox(0,0)[cc]{$M$}} \put(115,95){\makebox(0,0)[cc]{$N$}} \end{picture}
\end{array}
\]
For a scalar bosonic field $\varphi(x)$ the matrix element $\,\langle
\,\theta_{1}\,|\,\varphi(0)\,|\,\theta_{2},\theta_{3}\rangle$ is%
\begin{gather}
F^{\varphi}(\theta_{1};\theta_{2},\theta_{3})=F^{\varphi}(\theta_{1}+i\pi
_{-},\theta_{2},\theta_{3})+\delta_{\theta_{12}}+\delta_{\theta_{13}}%
S(\theta_{23})\label{cr}\\
\unitlength.25mm\begin{picture}(335,70)(10,0) \put(30,35){\oval(50,30)[]} \put(20,0){\line(0,1){20}} \put(40,0){\line(0,1){20}} \put(30,50){\line(0,1){20}} \put(30,35){\makebox(0,0)[cc]{$\varphi$}} \put(70,35){=} \put(20,-50){ \put(110,85){\oval(50,30)[]} \put(110,85){\makebox(0,0)[cc]{$\varphi$}} \put(85,70){\oval(20,30)[b]} \put(125,50){\line(0,1){20}} \put(110,50){\line(0,1){20}} \put(75,70){\line(0,1){50}} } \put(170,35){+} \put(140,-50){ \put(100,100){\oval(30,30)[]} \put(100,100){\makebox(0,0)[cc]{$\varphi$}} \put(100,50){\line(0,1){35}} \put(70,85){\oval(20,20)[]} \put(70,85){\makebox(0,0)[cc]{\bf 1}} \put(70,50){\line(0,1){25}} \put(70,95){\line(0,1){25}} } \put(270,35){+} \put(235,-50){ \put(100,100){\oval(30,30)[]} \put(100,100){\makebox(0,0)[cc]{$\varphi$}} \put(100,80){\line(0,1){5}} \put(70,95){\oval(20,20)[]} \put(70,95){\makebox(0,0)[cc]{\bf 1}} \put(70,80){\line(0,1){5}} \put(70,105){\line(0,1){15}} \put(70,50){\line(1,1){30}} \put(100,50){\line(-1,1){30}} } \end{picture}\nonumber
\end{gather}
and for $\,\langle\,\theta_{3},\theta_{2}\,|\,\varphi(0)\,|\,\theta_{4}%
\rangle$ we have
\begin{gather*}
F^{\varphi}(\theta_{2},\theta_{3};\theta_{4})=F^{\varphi}(\theta_{3}+i\pi
_{-},\theta_{2}+i\pi_{-},\theta_{4})+\delta_{\theta_{24}}+S(\theta_{32}%
)\delta_{\theta_{34}}\\
\unitlength.25mm\begin{picture}(345,70)(10,0) \put(30,35){\oval(50,30)[]} \put(20,50){\line(0,1){20}} \put(40,50){\line(0,1){20}} \put(30,0){\line(0,1){20}} \put(30,35){\makebox(0,0)[cc]{$\varphi$}} \put(70,35){=} \put(30,-50){ \put(110,85){\oval(50,30)[]} \put(110,85){\makebox(0,0)[cc]{$\varphi$}} \put(85,70){\oval(20,10)[b]} \put(85,70){\oval(50,30)[b]} \put(120,50){\line(0,1){20}} \put(75,70){\line(0,1){50}} \put(60,70){\line(0,1){50}} } \put(175,35){+} \put(140,-50){ \put(110,85){\oval(30,30)[]} \put(110,85){\makebox(0,0)[cc]{$\varphi$}} \put(100,70){\oval(20,30)[b]} \put(90,70){\line(0,1){50}} \put(70,85){\oval(20,20)[]} \put(70,85){\makebox(0,0)[cc]{\bf 1}} \put(70,50){\line(0,1){25}} \put(70,95){\line(0,1){25}} } \put(270,35){+} \put(235,-50){ \put(110,85){\oval(30,30)[]} \put(110,85){\makebox(0,0)[cc]{$\varphi$}} \put(100,70){\oval(20,30)[b]} \put(90,70){\line(0,1){30}} \put(70,85){\oval(20,20)[]} \put(70,85){\makebox(0,0)[cc]{\bf 1}} \put(70,50){\line(0,1){25}} \put(70,95){\line(0,1){5}} \put(70,100){\line(1,1){20}} \put(90,100){\line(-1,1){20}} } \end{picture}
\end{gather*}
with $i\pi_{\pm}=i\pi\pm i\epsilon$ and $\delta_{\theta_{12}}=4\pi
\delta(\theta_{1}-\theta_{2})$. Using the form factor equation (iii) and
Lorentz invariance (see e.g. \cite{BFK})
\begin{align*}
\operatorname*{Res}_{\theta_{12}=i\pi}F(\theta_{1},\theta_{2},\theta_{3})  &
=2i\,\left(  \mathbf{1}-S(\theta_{23})\right) \\
F(\theta_{1},\theta_{2},\theta_{3})  &  =F(\theta_{1}+\mu,\theta_{2}%
+\mu,\theta_{3}+\mu)
\end{align*}
we can rewrite these equations as (\ref{F12}) and (\ref{F21}). And further one
derives%
\begin{align}
&  \tfrac{1}{2}F(\theta_{1},\theta_{2}-i\pi_{-},\theta_{3}-i\pi_{+}%
)+\delta_{\theta_{12}}+\delta_{\theta_{13}}\label{F12a}\\
&  =\tfrac{1}{2}\left(  F(\theta_{1},\theta_{2}-i\pi_{+},\theta_{3}-i\pi
_{-})+\delta_{\theta_{12}}\left(  1+S(\theta_{23})\right)  +\delta
_{\theta_{13}}\left(  1+S(\theta_{23})\right)  \right) \nonumber\\
&  \tfrac{1}{2}F(\theta_{3}+i\pi_{+},\theta_{2}+i\pi_{-},\theta_{4}%
)+\delta_{\theta_{24}}+\delta_{\theta_{34}}\label{F21a}\\
&  =\tfrac{1}{2}\left(  F(\theta_{3}+i\pi_{-},\theta_{2}+i\pi_{+},\theta
_{4})+\delta_{\theta_{24}}\left(  1+S(\theta_{32})\right)  +\delta
_{\theta_{34}}\left(  1+S(\theta_{32})\right)  \right)  \,.\nonumber
\end{align}

\setcounter{equation}{0}

\section[Green's function and intermediate states]{Green's function and
intermediate states in low particle approximation}

\label{a1}

Let $\varphi(x)$ a scalar charge-less bosonic field with the
normalization\newline$\langle\,0\,|\,\varphi(x)\,|\,\theta\,\rangle=1$.

\paragraph{Simple examples of Wigthman and Green's functions}

\subparagraph{w1:}

The 2-point Wigthman function in 1-intermediate particle approximation is%
\[
w^{1}(x_{1}-x_{2})=\int_{\theta}\langle\,0\,|\,\varphi(x_{1})\,|\,\theta
\,\rangle\langle\,\theta\,|\,\varphi(x_{2})\,|\,0\rangle=i\Delta_{+}\left(
x_{1}-x_{2}\right)
\]
and the Green's function in this approximation is the free Feynman propagator%
\[
\tau^{1}(x)=\Theta(t)w^{1}(x)+\Theta(-t)w^{1}(-x)=\Delta_{F}\left(
x_{1}-x_{2}\right)  =\int_{p}e^{-ixp}\frac{i}{p^{2}-m^{2}+i\epsilon}%
\]

\subparagraph{w101:}

The 4-point Wightman function in 1-0-1-intermediate particle approximation is
(with $\int_{\theta}=\frac{1}{4\pi}\int d\theta$)%
\begin{align}
w^{101}(\underline{x})  &  =\int_{\theta_{1}}\langle\,0\,|\,\varphi
(x_{1})\,|\,\theta_{1}\,\rangle\langle\,\theta_{1}\,|\,\varphi(x_{2}%
)\,|\,0\rangle\,\int_{\theta_{4}}\langle\,0\,|\,\varphi(x_{3})\,|\,\theta
_{4}\rangle\langle\,\theta_{4}\,|\,\varphi(x_{4})\,|\,0\,\rangle\nonumber\\
&  =w^{1}(x_{1}-x_{2})w^{1}(x_{3}-x_{4})\label{w0}%
\end{align}
which implies that also%
\[
\tau^{101}(\underline{x})=\tau^{1}(x_{1}-x_{2})\tau^{1}(x_{3}-x_{4}).
\]

\subparagraph{w121:}

The 4-point Wightman function in 1-2-1-intermediate particle approximation is
(with $\int_{\underline{\theta}}=\int_{\theta_{1}}\dots\int_{\theta_{3}}$)%
\begin{align}
w^{121}(\underline{x})  &  =\frac{1}{2}\int_{\underline{\theta}}%
\langle\,0\,|\,\varphi(x_{1})\,|\,\theta_{1}\,\rangle\,\langle\,\theta
_{1}\,|\,\varphi(x_{2})\,|\,\theta_{2},\theta_{3}\rangle\,\langle\,\theta
_{3},\theta_{2}\,|\,\varphi(x_{3})\,|\,\theta_{4}\rangle\,\langle\,\theta
_{4}\,|\,\varphi(x_{4})\,|\,0\,\rangle\nonumber\\
&  =\frac{1}{2}\int_{\underline{\theta}}e^{-ix_{1}p_{1}-ix_{2}\left(
p_{2}+p_{3}-p_{1}\right)  -ix_{3}\left(  p_{4}-p_{2}-p_{3}\right)
+ix_{4}p_{4}}F(\theta_{1};\theta_{2},\theta_{3})F(\theta_{2},\theta_{3}%
;\theta_{4}).\label{w}%
\end{align}
Using equations (\ref{F12}, \ref{F21}) and the identity

$\left(  a+b+c\right)  \left(  d+e+f\right)  =\left(  \tfrac{1}{2}%
a+b+c\right)  d+a\left(  \tfrac{1}{2}d+e+f\right)  +\left(  b+c\right)
\left(  e+f\right)  $ we derive%
\begin{align*}
&  F(\theta_{1};\theta_{2},\theta_{3})F(\theta_{2},\theta_{3};\theta_{4})\\
&  =\left(  F(\theta_{1},\theta_{2}-i\pi_{-},\theta_{3}-i\pi_{+}%
)+\delta_{\theta_{12}}+\delta_{\theta_{13}}\right) \\
&  \times\left(  F(\theta_{3}+i\pi_{+},\theta_{2}+i\pi_{-},\theta_{4}%
)+\delta_{\theta_{24}}+\delta_{\theta_{34}}\right) \\
&  =\left(  \tfrac{1}{2}F(\theta_{1},\theta_{2}-i\pi_{-},\theta_{3}-i\pi
_{+})+\delta_{\theta_{12}}+\delta_{\theta_{13}}\right)  F(\theta_{3}+i\pi
_{+},\theta_{2}+i\pi_{-},\theta_{4})\\
&  +F(\theta_{1},\theta_{2}-i\pi_{-},\theta_{3}-i\pi_{+})\left(  \tfrac{1}%
{2}F(\theta_{3}+i\pi_{+},\theta_{2}+i\pi_{-},\theta_{4})+\delta_{\theta_{24}%
}+\delta_{\theta_{34}}\right) \\
&  +\left(  \delta_{\theta_{1}\theta_{2}}+\delta_{\theta_{1}\theta_{3}%
}\right)  \left(  \delta_{\theta_{4}\theta_{2}}+\delta_{\theta_{4}\theta_{3}%
}\right)
\end{align*}
which is using (\ref{F12a}) and (\ref{F21a}) equal to%
\begin{align*}
&  =\tfrac{1}{2}\left(  F(\theta_{1},\theta_{2}-i\pi_{+},\theta_{3}-i\pi
_{-})+\delta_{\theta_{12}}\left(  1+S(\theta_{23})\right)  +\delta
_{\theta_{13}}\left(  1+S(\theta_{23})\right)  \right) \\
&  \times F(\theta_{3}+i\pi_{+},\theta_{2}+i\pi_{-},\theta_{4})\\
&  +F(\theta_{1},\theta_{2}-i\pi_{-},\theta_{3}-i\pi_{+})\\
&  \times\tfrac{1}{2}\left(  F(\theta_{3}+i\pi_{-},\theta_{2}+i\pi_{+}%
,\theta_{4})+\delta_{\theta_{24}}\left(  1+S(\theta_{32})\right)
+\delta_{\theta_{34}}\left(  1+S(\theta_{32})\right)  \right) \\
&  +\left(  \delta_{\theta_{1}\theta_{2}}+\delta_{\theta_{1}\theta_{3}%
}\right)  \left(  \delta_{\theta_{4}\theta_{2}}+\delta_{\theta_{4}\theta_{3}%
}\right)
\end{align*}%
\begin{align}
&  =\tfrac{1}{2}F(\theta_{1},\theta_{2}-i\pi_{+},\theta_{3}-i\pi_{-}%
)F(\theta_{3}+i\pi_{+},\theta_{2}+i\pi_{-},\theta_{4})\label{I123}\\
&  +F(\theta_{1},\theta_{2}-i\pi_{-},\theta_{3}-i\pi_{+})\tfrac{1}{2}%
F(\theta_{3}+i\pi_{-},\theta_{2}+i\pi_{+},\theta_{4})\nonumber\\
&  +\tfrac{1}{2}\left(  \delta_{\theta_{12}}\left(  1+S(\theta_{23})\right)
+\delta_{\theta_{13}}\left(  1+S(\theta_{23})\right)  \right)  F(\theta
_{3}+i\pi_{+},\theta_{2}+i\pi_{-},\theta_{4})\nonumber\\
&  +F(\theta_{1},\theta_{2}-i\pi_{-},\theta_{3}-i\pi_{+})\tfrac{1}{2}\left(
\delta_{\theta_{24}}\left(  1+S(\theta_{32})\right)  +\delta_{\theta_{34}%
}\left(  1+S(\theta_{32})\right)  \right) \nonumber\\
&  +\left(  \delta_{\theta_{1}\theta_{2}}+\delta_{\theta_{1}\theta_{3}%
}\right)  \left(  \delta_{\theta_{4}\theta_{2}}+\delta_{\theta_{4}\theta_{3}%
}\right) \nonumber\\
&  =I_{1}(\underline{\theta})+I_{2}(\underline{\theta})+I_{3}(\underline
{\theta})\nonumber
\end{align}
where we have introduced%
\begin{align}
I_{1}(\underline{\theta})  &  =\tfrac{1}{2}F(\theta_{1},\theta_{2}-i\pi
_{+},\theta_{3}-i\pi_{-})F(\theta_{3}+i\pi_{+},\theta_{2}+i\pi_{-},\theta
_{4})\nonumber\\
&  +\tfrac{1}{2}F(\theta_{1},\theta_{2}-i\pi_{-},\theta_{3}-i\pi_{+}%
)F(\theta_{3}+i\pi_{-},\theta_{2}+i\pi_{+},\theta_{4})\nonumber\\
I_{2}(\underline{\theta})  &  =\tfrac{1}{2}\left(  \delta_{\theta_{12}}\left(
1+S(\theta_{23})\right)  +\delta_{\theta_{13}}\left(  1+S(\theta_{23})\right)
\right)  F(\theta_{3}+i\pi_{+},\theta_{2}+i\pi_{-},\theta_{4})\label{I}\\
&  +\tfrac{1}{2}F(\theta_{1},\theta_{2}-i\pi_{-},\theta_{3}-i\pi_{+})\left(
\delta_{\theta_{24}}\left(  1+S(\theta_{32})\right)  +\delta_{\theta_{34}%
}\left(  1+S(\theta_{32})\right)  \right) \nonumber\\
I_{3}(\underline{\theta})  &  =\left(  \delta_{\theta_{12}}+\delta
_{\theta_{13}}\right)  \left(  \delta_{\theta_{24}}+\delta_{\theta_{34}%
}\right) \nonumber
\end{align}
From $I_{3}$ we calculate%
\begin{equation}
w_{3}^{121}(\underline{x})=w^{1}\left(  x_{1}-x_{4}\right)  w^{1}\left(
x_{2}-x_{3}\right)  +w^{1}\left(  x_{1}-x_{3}\right)  w^{1}\left(  x_{2}%
-x_{4}\right)  \,\label{w3}%
\end{equation}
which together with (\ref{w0}) yields the disconnected part of $\tilde{\tau
}(\underline{k})=\tilde{\tau}_{disc}(\underline{k})+\tilde{\tau}%
_{c}(\underline{k})$. This means the connected part of the Green's function is
given by $I_{1}(\underline{\theta})+I_{2}(\underline{\theta})$.

\paragraph{General case}

We start with (\ref{tau}), insert sets of intermediate states and write
$y_{i}=x_{\pi i}$%
\begin{align*}
\tilde{\tau}_{\underline{\varphi}}(\underline{k})  &  =\sum_{\pi\in S_{n}}%
\sum\frac{1}{\underline{n}!}\int\underline{d^{2}y}e^{iy_{i}k_{\pi i}}%
\Theta_{1\dots n}(\underline{y})\\
&  \times\int_{\underline{p^{(1)}}}\dots\int_{\underline{p^{(n-1)}}}%
\langle0|\varphi_{\pi1}(y_{1})|\underline{p^{(1)}}\rangle\langle
\underline{p^{(1)}}|\dots|\underline{p^{(n-1)}}\rangle\langle\underline
{p^{(n-1)}}|\,\varphi_{\pi n}(y_{n})|0\rangle
\end{align*}
with the notation of (\ref{Xin}). We perform the $y$-integrations and obtain
for $\tilde{\Xi}$ defined in (\ref{Xi})%
\begin{multline*}
\tilde{\Xi}_{\underline{\varphi}}(\underline{k})=\sum_{\pi\in S_{n}}\sum
\frac{1}{\underline{n}!}\int_{\underline{p^{(1)}}}\dots\int_{\underline
{p^{(n-1)}}}\langle0|\varphi_{\pi1}(0)|\underline{p^{(1)}}\rangle
\langle\underline{p^{(1)}}|\dots\underline{|p^{(n-1)}}\rangle\langle
\underline{p^{(n-1)}}|\,\varphi_{\pi n}(0)|0\rangle\\
\times2\pi\delta\left(  k_{\pi2}-\sum p_{j}^{(2)}+\sum p_{j}^{(1)}\right)
^{1}\dots2\pi\delta\left(  k_{\pi n}+\sum p_{j}^{(n-1)}\right)  ^{1}\\
\times\frac{-i}{\sum_{i=2}^{n}k_{\pi i}^{0}+\sum\omega_{j}^{(1)}-i\epsilon
}\frac{-i}{\sum_{i=3}^{n}k_{\pi i}^{0}+\sum\omega_{j}^{(2)}-i\epsilon}%
\dots\frac{-i}{k_{\pi n}^{0}+\sum\omega_{j}^{(n-1)}-i\epsilon}%
\end{multline*}
because%
\begin{multline*}
\int\underline{dy^{1}}e^{-iy_{i}^{1}\left(  k_{\pi i}-\sum p_{j}^{(i)}+\sum
p_{j}^{(i-1)}\right)  ^{1}}\\
=2\pi\delta\left(  \sum k_{i}^{1}\right)  2\pi\delta\left(  k_{\pi2}-\sum
p_{j}^{(2)}+\sum p_{j}^{(1)}\right)  ^{1}\dots2\pi\delta\left(  k_{\pi n}+\sum
p_{j}^{(n-1)}\right)  ^{1}%
\end{multline*}
and%
\begin{multline*}
\int\underline{dy^{0}}\Theta_{1\dots n}(\underline{y})e^{iy_{i}^{0}\left(
k_{\pi i}-\sum p_{j}^{(i)}+\sum p_{j}^{(i-1)}\right)  ^{0}}=2\pi\delta\left(
\sum k_{i}^{0}\right) \\
\times\frac{-i}{k_{\pi2}^{0}+\dots+k_{\pi n}^{0}+\sum\omega_{j}^{(1)}%
-i\epsilon}\frac{-i}{k_{\pi3}^{0}+\dots+k_{\pi n}^{0}+\sum\omega_{j}%
^{(2)}-i\epsilon}\dots\frac{-i}{k_{\pi n}^{0}+\sum\omega_{j}^{(n-1)}%
-i\epsilon}%
\end{multline*}
which proves (\ref{Xin}). For $k_{i}^{1}\rightarrow0$ the result is%
\begin{align}
\tilde{\Xi}_{\underline{\varphi}}(\underline{k})  &  =\sum_{\pi\in S_{n}}%
\sum\frac{1}{\underline{n}!}\int_{\underline{p^{(1)}}}\dots\int_{\underline
{p^{(n-1)}}}\langle\varphi_{\pi1}(0)|\underline{p^{(1)}}\rangle\langle
\underline{p^{(1)}}|\dots\underline{|p^{(n-1)}}\rangle\langle\underline
{p^{(n-1)}}|\,\varphi_{\pi n}(0)\rangle\label{Xigen0}\\
&  \times2\pi\delta\left(  \sum p_{j}^{(1)}\right)  ^{1}\dots2\pi\delta\left(
\sum p_{j}^{(n-1)}\right)  ^{1}\nonumber\\
&  \times\frac{-i}{\sum_{i=2}^{n}k_{\pi i}^{0}+\sum\omega_{j}^{(1)}-i\epsilon
}\frac{-i}{\sum_{i=3}^{n}k_{\pi i}^{0}+\sum\omega_{j}^{(2)}-i\epsilon}%
\dots\frac{-i}{k_{\pi n}^{0}+\sum\omega_{j}^{(n-1)}-i\epsilon}\nonumber
\end{align}

\paragraph{Example: the 4-point function}

Let $\underline{\varphi}=\varphi\varphi\varphi\varphi$. We use the
intermediate states

$\left\langle 0|\varphi(0)|p_{1}\rangle\langle p_{1}|\varphi(0)|p_{2}%
,p_{3}\rangle\langle p_{3},p_{2}|\varphi(0)|p_{4}\rangle\langle p_{4}%
|\,\varphi(0)|0\right\rangle $ and as in (\ref{w}) we obtain in this
approximation%
\begin{align*}
\tilde{\Xi}_{\underline{\varphi}}(\underline{k})  &  =\sum_{\pi\in S_{n}}%
\int_{p_{1}}\int_{p_{2}}\int_{p_{3}}\int_{p_{4}}2\pi\delta\left(
p_{1}\right)  2\pi\delta\left(  p_{2}+p_{3}\right)  2\pi\delta\left(
p_{4}\right) \\
&  \times\frac{1}{2!}\left\langle 0|\varphi(0)|p_{1}\rangle\langle
p_{1}|\varphi(0)|p_{2},p_{3}\rangle\langle p_{3},p_{2}|\varphi(0)|p_{4}%
\rangle\langle p_{4}|\,\varphi(0)|0\right\rangle \\
&  \times\frac{-i}{-k_{\pi1}^{0}+\omega_{1}-i\epsilon}\frac{-i}{k_{\pi3}%
^{0}+k_{\pi4}^{0}+\omega_{2}+\omega_{3}-i\epsilon}\frac{-i}{k_{\pi4}%
^{0}+\omega_{4}-i\epsilon}\\
&  =-\frac{1}{32}\frac{i}{m^{6}\pi}\sum_{perm(k)}\frac{m}{-k_{1}%
^{0}+m-i\epsilon}\frac{m}{k_{4}^{0}+m-i\epsilon}g\left(  -\left(  k_{3}%
^{0}+k_{4}^{0}\right)  /(2m)+i\epsilon\right) \\
g(x)  &  =\frac{-1}{4}\int d\theta\frac{1}{\cosh\theta}\frac{1}{\cosh\theta
-x}F^{\varphi}(0;\theta,-\theta)F^{\varphi}(\theta,-\theta;0)
\end{align*}
which is (\ref{Sigma4}) and $F(0;\theta,-\theta)F(\theta,-\theta;0)$ is given
by (see (\ref{I}))
\[
F(\theta_{1};\theta_{2},\theta_{3})F(\theta_{2},\theta_{3};\theta_{4}%
)=I_{1}(\underline{\theta})+I_{2}(\underline{\theta})+I_{3}(\underline{\theta
}).
\]
As mentioned in the context of (\ref{w3}) the connected part of $\tilde{\Xi
}_{\underline{\varphi}}(\underline{k})$ is obtained by
\[
g_{1}(x)+g_{2}(x)=\frac{-1}{4}\int d\theta\frac{1}{\cosh\theta}\frac{1}%
{\cosh\theta-x}\left(  I_{1}(\underline{\theta})+I_{2}(\underline{\theta
})\right)  \,.
\]
The contribution from $I_{1}$ will be calculated for the $Z(2)$-, the $Z(3)$-
scaling Ising and the sinh-Gordon models. The contribution from $I_{2}$ leads
to $0/0$, therefore the limit $k_{i}^{1}\rightarrow0$ has to be taken more carefully.

\subparagraph{Contribution of $I_{2}$:}

For $k_{i}=(k_{i}^{0},m\sinh\kappa_{i})$ this is equal to%
\begin{align*}
\tilde{\Xi}_{\underline{\varphi}2}(\underline{k})  &  =\frac{1}{2}%
\sum_{perm(k)}\frac{1}{\left(  2m\right)  ^{2}}\int_{p_{2}}\int_{p_{3}}\left(
2\pi\right)  \delta(k_{1}^{1}+k_{2}^{1}-p_{2}^{1}-p_{3}^{1})I_{2}(\kappa
_{1},\theta_{2},\theta_{3},-\kappa_{4})\\
&  \times\frac{-i}{-k_{1}^{0}+\omega_{1}-i\epsilon}\frac{-i}{k_{3}^{0}%
+k_{4}^{0}+\omega_{2}+\omega_{3}-i\epsilon}\frac{-i}{k_{4}^{0}+\omega
_{4}-i\epsilon}%
\end{align*}
with $I_{2}(\kappa_{1},\theta_{2},\theta_{3},-\kappa_{4})$ given by (\ref{I}).
Taking first the term with $\delta_{\kappa_{1}\theta_{2}}$ we get (because
$\theta_{3}\rightarrow\kappa_{2}$)%
\begin{align*}
&  \frac{1}{2}\frac{1}{\left(  2m\right)  ^{2}}\sum_{perm(k)}\frac{1}%
{2m\cosh\kappa_{2}}\tfrac{1}{2}\left(  1+S(\kappa_{12})\right)  F^{\varphi
}(\kappa_{2}+i\pi_{+},\kappa_{1}+i\pi_{-},-\kappa_{4})\\
&  \times\frac{-i}{-k_{1}^{0}+m\cosh\kappa_{1}}\frac{-i}{k_{3}^{0}+k_{4}%
^{0}+m\cosh\kappa_{1}+m\cosh\kappa_{2}}\frac{-i}{k_{4}^{0}+m\cosh\kappa_{4}}.
\end{align*}
We write%
\[
F^{\varphi}(\theta_{1},\theta_{2},\theta_{3})=\prod_{i<j}\tanh\tfrac{1}%
{2}\theta_{ij}~\tilde{F}^{\varphi}(\theta_{1},\theta_{2},\theta_{3})
\]
then for small $\kappa_{i}$ (using $S(0)=-1$)%
\begin{align*}
&  \tfrac{1}{2}\left(  1+S(\kappa_{12})\right)  F^{\varphi}(\kappa_{2}%
+i\pi_{+},\kappa_{1}+i\pi_{-},-\kappa_{4})\\
&  \rightarrow\tfrac{1}{2}\kappa_{12}S^{\prime}(0)\tanh\frac{1}{2}\kappa
_{21}\coth\frac{1}{2}\left(  \kappa_{2}+\kappa_{4}\right)  \coth\frac{1}%
{2}\left(  \kappa_{1}+\kappa_{4}\right)  \tilde{F}^{\varphi}(i\pi,i\pi,0)\\
&  \rightarrow\frac{-\kappa_{21}^{2}}{\left(  \kappa_{2}+\kappa_{4}\right)
\left(  \kappa_{1}+\kappa_{4}\right)  }S^{\prime}(0)\tilde{F}^{\varphi}%
(i\pi,i\pi,0)
\end{align*}
Similarly we get the other contribution from $I_{2}$ and calculate%
\begin{multline*}
\frac{\left(  \kappa_{1}-\kappa_{2}\right)  ^{2}}{(\kappa_{2}+\kappa
_{4})(\kappa_{1}+\kappa_{4})}+\frac{\left(  \kappa_{1}-\kappa_{3}\right)
^{2}}{(\kappa_{3}+\kappa_{4})(\kappa_{1}+\kappa_{4})}\\
+\frac{\left(  \kappa_{3}-\kappa_{4}\right)  ^{2}}{\left(  \kappa_{1}%
+\kappa_{4}\right)  \left(  \kappa_{1}+\kappa_{3}\right)  }+\frac{\left(
\kappa_{2}-\kappa_{4}\right)  ^{2}}{\left(  \kappa_{1}+\kappa_{4}\right)
\left(  \kappa_{1}+\kappa_{2}\right)  }=-8
\end{multline*}
up to terms proportional to $\kappa_{1}+\kappa_{2}+\kappa_{3}+\kappa_{4}$,
which do not contribute because of the $\delta$-function $\delta\left(
k_{1}^{1}+k_{2}^{1}+k_{3}^{1}+k_{4}^{1}\right)  $. Therefore in the limit
$\kappa_{i}\rightarrow0$ %
\[
\tilde{\Xi}_{\underline{\varphi}2}(\underline{k})=-\frac{1}{32}\frac{i}%
{m^{6}\pi}\sum_{perm(k)}\frac{m}{-k_{1}^{0}+m-i\epsilon}\frac{m}{k_{4}%
^{0}+m-i\epsilon}g_{2}\left(  -\left(  k_{3}^{0}+k_{4}^{0}\right)
/(2m)+i\epsilon\right)
\]%
\begin{equation}
g_{2}(x)=-8\pi S^{\prime}(0)\tilde{F}^{\varphi}(i\pi,i\pi,0)\frac{1}%
{1-x}.\label{g2}%
\end{equation}

\setcounter{equation}{0}

\section{Models}

\subsection{The scaling $Z_{2}$ Ising model}

\label{aZ2}

The Quantum Ising model is described by the Hamiltonian
\begin{equation}
H=\sum_{n}\Big(-J\sigma_{n}^{z}\sigma_{n+1}^{z}+h\sigma_{n}^{x}%
\Big),\label{modelI}%
\end{equation}
where $\sigma^{a}$ are the Pauli matrices. This model has numerous condensed
matter realizations being one of the most popular models of condensed matter
theory. It describes a sequence of coupled two level systems. They may
represent spins; then the first term describes an anisotropic exchange
interaction. In this case $\sigma^{z}$ directly couples to external magnetic
field: $\mu_{B}B_{n}^{z}\sigma_{n}^{z}$.

States of the two level systems may also correspond to positions of electric
charges in a double well potential. Then the first term is the dipole-dipole
interaction and the transverse field describes the quantum tunnelling between
the wells. Then $\sigma^{a}$ would be the dipole moment operators. Their
interaction with the electric field is given by $pE_{n}^{z}\sigma_{n}^{z}$
with $p$ being the dipole moment.

Since the dominant interaction is ferromagnetic, the strongest fluctuations
take place at zero wave vectors which guarantees a direct coupling to the
electromagnetic field creating optimal resonance conditions. The Ising model
(\ref{modelI}) has two phases depending on the sign of $m=h-J$. The resonance
occurs in the paramagnetic phase $m>0$ when the ground state average of the
order parameter $\langle\,\sigma^{z}\,\rangle=0$. In that case the
electromagnetic field has a nonzero matrix element between the ground state
and single magnon state.

In the scaling limit model (\ref{modelI}) can be described by an interacting
Bose field $\sigma_{n}^{z}=Cm^{1/8}\sigma(x)$, where $C$ is a numerical
constant and $m=h-J$. The excitations are non-interacting Majorana fermions
with the 2-particle S-matrix $S^{Z(2)}(\theta)=-1$. The field $\sigma
^{x}=(m/J)^{1/2}\epsilon(x)\sim\bar{\psi}\psi(x)$, where $\psi$ is a free
Majorana spinor field. The $n$-particle form factors for the order parameter
$\sigma(x)$ is given by (\ref{Fn}). From $\epsilon(x)\sim\bar{\psi}\psi(x)$
one has for a free Majorana spinor field (up to a constant)
\begin{equation}
\langle0|\epsilon(0)|\theta_{1},\theta_{2}\rangle=\sinh\tfrac{1}{2}\theta
_{12}\,.\label{e2}%
\end{equation}

\subsubsection{The 3-point function}

We calculate $\tilde{\Xi}_{\varphi\varphi\epsilon}(\underline{k})$ in the
limit $k_{i}^{1}\rightarrow0$. For the various permutations in (\ref{Xin}) we
obtain:\newline a) For the permutation $\pi=123$ and $n_{1}=1,~n_{2}=2$%

\begin{align*}
&  \tilde{\Xi}_{\varphi\varphi\epsilon}^{12}(k_{1},k_{2},k_{3})=\frac{1}%
{2!}\int_{p_{1}}\int_{p_{2}}\int_{p_{3}}2\pi\delta\left(  p_{1}\right)
2\pi\delta\left(  p_{2}+p_{3}\right)  \langle0|\varphi(0)|p_{1}\rangle\\
&  \times\langle p_{1}|\varphi(0)|p_{2},p_{3}\rangle\langle p_{3}%
,p_{2}|\epsilon(0)|0\rangle\frac{-i}{-k_{1}^{0}+m-i\epsilon}\frac{-i}%
{k_{3}^{0}+\omega_{2}+\omega_{3}-i\epsilon}\\
&  =-\frac{1}{64m^{4}\pi}\frac{m}{-k_{1}^{0}+m-i\epsilon}\int d\theta
\frac{2mF^{\varphi}(i\pi,\theta,-\theta)F^{\epsilon}(-\theta+i\pi,\theta
+i\pi)}{k_{3}^{0}+2\omega-i\epsilon}%
\end{align*}
which is (\ref{Xi12}). Equations (\ref{Fn}) and (\ref{e2}) imply%
\begin{multline*}
F^{\varphi}(i\pi,\theta,-\theta)F^{\epsilon}(-\theta+i\pi,\theta+i\pi)\\
=2i\tanh\tfrac{1}{2}\left(  i\pi-\theta\right)  \,\tanh\tfrac{1}{2}\left(
i\pi+\theta\right)  \,\tanh\theta\,\sinh\left(  -\theta\right)  =2i\frac
{\left(  \cosh\theta+1\right)  ^{2}}{\cosh\theta}%
\end{multline*}
and therefore%
\begin{align*}
\tilde{\Xi}_{\varphi\varphi\epsilon}^{12}(k_{1},k_{2},k_{3})  &  =-\frac
{i}{32\pi m^{4}}\frac{m}{-k_{1}^{0}+m-i\epsilon}h_{+}^{Z(2)}(-k_{3}%
^{0}/(2m)+i\epsilon)\\
h_{+}^{Z(2)}(x)  &  =\int_{-\infty}^{\infty}d\theta\frac{\left(  \cosh
\theta+1\right)  ^{2}}{\cosh^{2}\theta}\frac{1}{\cosh\theta-x}\\
&  =-\frac{2}{x}-\frac{2}{x}\pi-\frac{1}{x^{2}}\pi-4\frac{\left(  1+x\right)
^{2}}{x^{2}\sqrt{x^{2}-1}}\operatorname{arctanh}\frac{1+x}{\sqrt{x^{2}-1}}\,.
\end{align*}
b) For the permutation $\pi=321$ and $n_{1}=2,~n_{2}=1$%

\begin{multline*}
\tilde{\Xi}_{\epsilon\varphi\varphi}^{21}(k_{3},k_{2},k_{1})=\frac{1}{2!}%
\int_{p_{1}}\int_{p_{2}}\int_{p_{3}}2\pi\delta\left(  p_{1}+p_{2}\right)
2\pi\delta\left(  p_{3}\right)  \langle0|\epsilon(0)|p_{1},p_{2}\rangle\langle
p_{2},p_{1}|\varphi(0)|p_{3}\rangle\\
\times\langle p_{3}|\varphi(0)|0\rangle\frac{-i}{k_{3}^{0}+k_{2}^{0}%
+\omega_{1}+\omega_{2}-i\epsilon}\frac{-i}{k_{1}^{0}+\omega_{3}-i\epsilon}\\
=\frac{1}{2!}\frac{-i}{k_{1}^{0}+m-i\epsilon}\frac{1}{2m}\int_{\theta}\frac
{1}{2\omega}F^{\epsilon}\left(  \theta,-\theta\right)  F^{\varphi}%
(-\theta+i\pi,\theta+i\pi,0)\frac{-i}{-k_{3}^{0}+2\omega-i\epsilon}\\
=-\frac{1}{64m^{4}\pi}\frac{m}{k_{1}^{0}+m-i\epsilon}\int d\theta\frac
{m}{\omega}\frac{2mF^{\epsilon}\left(  \theta,-\theta\right)  F^{\varphi
}(-\theta+i\pi,\theta+i\pi,0)}{-k_{3}^{0}+2\omega-i\epsilon}%
\end{multline*}
which is (\ref{Xi21}) and
\[
F^{\epsilon}\left(  \theta,-\theta\right)  F^{\varphi}(-\theta+i\pi
,\theta+i\pi,0)=2i\frac{\left(  \cosh\theta+1\right)  ^{2}}{\cosh\theta}%
\]
imply again%
\[
\tilde{\Xi}_{\epsilon\varphi\varphi}^{21}(k_{3},k_{2},k_{1})=-\frac{i}{32\pi
m^{4}}\frac{m}{k_{1}^{0}+m-i\epsilon}h_{+}^{Z(2)}(k_{3}^{0}/(2m)+i\epsilon
)\,.\,
\]
c) For the permutation $\pi=132$ and $n_{1}=1,~n_{2}=1$%
\begin{align*}
\tilde{\Xi}_{\varphi\epsilon\varphi}^{11}(k_{1},k_{3},k_{2})  &  =\int_{p_{1}%
}\int_{p_{2}}2\pi\delta\left(  p_{1}\right)  2\pi\delta\left(  p_{2}\right)
\left\langle 0|\varphi(0)|p_{1}\rangle\langle p_{1}|\epsilon(0)|p_{2}%
\rangle\langle p_{2}|\varphi(0)|0\right\rangle \\
&  \times\frac{-i}{k_{3}^{0}+k_{2}^{0}+\omega_{1}-i\epsilon}\frac{-i}%
{k_{2}^{0}+\omega_{2}-i\epsilon}\\
&  =-\frac{1}{4m^{2}}\frac{m}{-k_{1}^{0}+m-i\epsilon}\frac{m}{k_{2}%
^{0}+m-i\epsilon}F^{\epsilon}\left(  i\pi,0\right) \\
&  =-\frac{i}{4m^{2}}\frac{m}{-k_{1}^{0}+m-i\epsilon}\frac{m}{k_{2}%
^{0}+m-i\epsilon}%
\end{align*}
which is (\ref{Xi11}) because $F^{\epsilon}\left(  i\pi,0\right)  =\sinh
\frac{1}{2}i\pi=i$.\newline d) For the permutation $\pi=132$ and
$n_{1}=3,~n_{2}=1$%

\begin{align*}
\tilde{\Xi}_{\varphi\epsilon\varphi}^{31}(k_{1},k_{3},k_{2})  &  =\frac{1}%
{3!}\int_{p_{1}}\int_{p_{2}}\int_{p_{3}}\int_{p_{4}}\langle0|\varphi
(0)|p_{1},p_{2},p_{3}\rangle\langle p_{3},p_{2},p_{1}|\epsilon(0)|p_{4}%
\rangle\langle p_{4}|\,\varphi(0)|0\rangle\\
&  \times2\pi\delta\left(  p_{1}^{1}+p_{2}^{1}+p_{3}^{1}\right)  2\pi
\delta\left(  p_{4}^{1}\right)  \frac{-i}{-k_{1}^{0}+\omega_{1}+\omega
_{2}+\omega_{3}-i\epsilon}\frac{-i}{k_{3}^{0}+\omega_{4}-i\epsilon}\\
&  =\frac{1}{3!}\frac{1}{2m}\int_{p_{1}}\int_{p_{2}}\int_{p_{3}}\frac
{-i}{-k_{1}^{0}+\omega_{1}+\omega_{2}+\omega_{3}-i\epsilon}\frac{-i}{k_{3}%
^{0}+m-i\epsilon}\\
&  \times2\pi\delta\left(  p_{1}^{1}+p_{2}^{1}+p_{3}^{1}\right)  F^{\varphi
}(\theta_{1},\theta_{2},\theta_{3})F^{\epsilon}(\theta_{1},\theta_{2}%
,\theta_{3};\theta_{4})
\end{align*}
There are 3 contributions from%
\begin{multline*}
F^{\epsilon}(\theta_{1},\theta_{2},\theta_{3};\theta_{4})=\delta_{\theta
_{1}\theta_{4}}F^{\epsilon}(\theta_{3}+i\pi,\theta_{2}+i\pi)\\
-\delta_{\theta_{2}\theta_{4}}F^{\epsilon}(\theta_{3}+i\pi,\theta_{1}%
+i\pi)+\delta_{\theta_{3}\theta_{4}}F^{\epsilon}(\theta_{2}+i\pi,\theta
_{1}+i\pi)
\end{multline*}
It turns out that all 3 give the same result, therefore%
\begin{align*}
\tilde{\Xi}_{\varphi\epsilon\varphi}^{31}(k_{1},k_{3},k_{2})  &  =3\frac
{1}{3!}\frac{1}{2m}\frac{-i}{k_{3}^{0}+m-i\epsilon}\int_{p_{1}}\int_{p_{2}%
}\int_{p_{3}}\frac{-i}{-k_{1}^{0}+\omega_{1}+\omega_{2}+\omega_{3}-i\epsilon
}\\
&  \times2\pi\delta\left(  p_{1}^{1}+p_{2}^{1}+p_{3}^{1}\right)  F(\theta
_{1},\theta_{2},\theta_{3})\delta_{\theta_{1}\theta_{4}}F^{\epsilon}%
(\theta_{3}+i\pi,\theta_{2}+i\pi)\\
&  =\frac{1}{2}\frac{\left(  -i\right)  ^{2}}{\left(  2m\right)  ^{3}m}%
\frac{1}{4\pi}\frac{m}{k_{3}^{0}+m-i\epsilon}\int d\theta\frac{2m}{2\omega
}\frac{2mF(0,\theta,-\theta)F^{\epsilon}(-\theta+i\pi,\theta+i\pi)}{-k_{1}%
^{0}+m+2\omega-i\epsilon}\\
&  =-\frac{1}{64m^{4}\pi}\frac{m}{k_{3}^{0}+m-i\epsilon}\int d\theta\frac
{m}{\omega}\frac{2mF(0,\theta,-\theta)F^{\epsilon}(-\theta+i\pi,\theta+i\pi
)}{-k_{1}^{0}+m+2\omega-i\epsilon}%
\end{align*}
and
\[
F^{\varphi}(0,\theta,-\theta)F^{\epsilon}(-\theta+i\pi,\theta+i\pi
)=2i\frac{\left(  \cosh\theta-1\right)  ^{2}}{\cosh\theta}%
\]%
\begin{align*}
\tilde{\Xi}_{\varphi\epsilon\varphi}^{31}(k_{1},k_{3},k_{2})  &  =\frac
{-1}{32m^{4}\pi}\frac{m}{k_{3}^{0}+m-i\epsilon}h_{-}^{Z(2)}(k_{1}%
^{0}/(2m)-\tfrac{1}{2})\\
h_{-}^{Z(2)}(x)  &  =\int_{-\infty}^{\infty}d\theta\frac{\left(  \cosh
\theta-1\right)  ^{2}}{\cosh^{2}\theta}\frac{1}{\cosh\theta-x}\\
&  =-\frac{2}{x}+\frac{2}{x}\pi-\frac{1}{x^{2}}\pi-4\frac{\left(  x-1\right)
^{2}}{x^{2}\sqrt{x^{2}-1}}\operatorname{arctanh}\frac{1+x}{\sqrt{x^{2}-1}}%
\end{align*}

d) For the permutation $\pi=132$ and $n_{1}=1,~n_{2}=3$ we find, similarly%
\[
\tilde{\Xi}_{\varphi\epsilon\varphi}^{13}(k_{1},k_{3},k_{2})=\frac{-1}%
{32m^{4}\pi}\frac{m}{-k_{1}^{0}+m-i\epsilon}h_{-}^{Z(2)}(-k_{2}^{0}%
/(2m)-\tfrac{1}{2}+i\epsilon).
\]
Finally with (\ref{Xi3}) we obtain (\ref{Xippe}).

\subsubsection{The 4-point function}

From (\ref{Sigma4}) for $k_{i}=(k_{i}^{0},0)$ in momentum space the
contribution from $I_{2}$ in (\ref{I}) vanishes, because $S(0)=-1$ and we get%
\begin{align*}
\tilde{\Xi}_{\underline{\varphi}}(\underline{k})  &  =-\frac{i}{32\pi m^{6}%
}\sum_{perm(k)}\frac{m}{-k_{1}^{0}+m-i\epsilon}\frac{m}{k_{4}^{0}+m-i\epsilon
}g^{Z2}\left(  -\frac{k_{3}^{0}+k_{4}^{0}}{2m}+i\epsilon\right) \\
g^{Z2}(x)  &  =-\frac{1}{4}\int d\theta\frac{1}{\cosh\theta}\frac{1}%
{\cosh\theta-x}I_{\underline{\varphi}}^{Z_{2}}(0,\theta,-\theta,0)
\end{align*}
From (\ref{I}) and (\ref{Fn}) we obtain%
\begin{align*}
&  I_{\underline{\varphi}}^{Z_{2}}(0,\theta,-\theta,0)\\
&  =\tfrac{1}{2}F(0,\theta-i\pi_{+},-\theta-i\pi_{-})F(-\theta+i\pi_{+}%
,\theta+i\pi_{-},0)+\left(  \epsilon\rightarrow-\epsilon\right) \\
&  =\tfrac{1}{2}\left(  2i\right)  ^{2}\tanh\frac{1}{2}\left(  -\theta
+i\pi+i\epsilon\right)  \tanh\frac{1}{2}\left(  \theta+i\pi-i\epsilon\right)
\tanh\frac{1}{2}\left(  2\theta\right) \\
&  \times\tanh\frac{1}{2}\left(  -2\theta\right)  \tanh\frac{1}{2}\left(
-\theta+i\pi+i\epsilon\right)  \tanh\frac{1}{2}\left(  \theta+i\pi
-i\epsilon\right)  +\left(  \epsilon\rightarrow-\epsilon\right) \\
&  =2\tanh^{2}\theta\coth^{4}\tfrac{1}{2}\left(  \theta-i\epsilon\right)
+\left(  \epsilon\rightarrow-\epsilon\right)  .
\end{align*}
and%
\begin{align*}
g^{Z(2)}(x)  &  =-\frac{1}{2}\int_{-\infty}^{\infty}d\theta\left(  \frac
{\coth^{4}\tfrac{1}{2}\left(  \theta-i\epsilon\right)  \,\tanh^{2}%
\theta+\left(  \epsilon\rightarrow-\epsilon\right)  }{\cosh\theta\,\left(
\cosh\theta-x\right)  }\right) \\
&  =-\int_{-\infty}^{\infty}\left(  \frac{\coth^{4}\tfrac{1}{2}\theta
\,\tanh^{2}\theta}{\cosh\theta}\frac{1}{\cosh\theta-x}-\frac{16}{\theta^{2}%
}\frac{1}{1-x}\right)  d\theta
\end{align*}
which can be calculated:\newline for $\operatorname{Re}x<-1$%
\[
g^{Z(2)}(x)=\frac{16}{1-x}-\frac{15\pi}{2x}-\frac{8}{x}-\frac{4\pi+2}{x^{2}%
}-\frac{\pi}{x^{3}}-\frac{\left(  x+1\right)  ^{2}\sqrt{x^{2}-1}}{x^{3}\left(
x-1\right)  ^{2}}2\ln\left(  -x+\sqrt{x^{2}-1}\right)
\]
for $\operatorname{Re}x>1$%
\begin{align*}
g^{Z(2)}(x\pm i\epsilon)  &  =\frac{16}{1-x}-\frac{15\pi}{2x}-\frac{8}%
{x}-\frac{4\pi+2}{x^{2}}-\frac{\pi}{x^{3}}\\
&  -\frac{\left(  x+1\right)  ^{2}\sqrt{x^{2}-1}}{x^{3}\left(  x-1\right)
^{2}}2\left(  \pm i\pi+\ln\left(  x+\sqrt{x^{2}-1}\right)  \right) \\
\operatorname{Im}g^{Z(2)}(x\pm i\epsilon)  &  =\mp\Theta(x-1)2\pi\frac{\left(
x+1\right)  ^{2}\sqrt{x^{2}-1}}{x^{3}\left(  x-1\right)  ^{2}}%
\end{align*}
for $-1<x<1$%
\begin{align*}
g^{Z(2)}(x)  &  =\frac{16}{1-x}-\frac{15\pi}{2x}-\frac{8}{x}-\frac{4\pi
+2}{x^{2}}-\frac{\pi}{x^{3}}-\frac{\left(  x+1\right)  ^{2}i\sqrt{1-x^{2}}%
}{x^{3}\left(  x-1\right)  ^{2}}2\ln\left(  -x+i\sqrt{1-x^{2}}\right) \\
&  =\left(  \frac{94}{3}+10\pi\right)  +O\left(  x\right)
\end{align*}
The intrinsic coupling $g_{R}$ \cite{BNNPSW}, defined by $\tilde{\Xi
}(0)=-\frac{i}{m^{6}}g_{R}$ is
\[
g_{R}=\frac{47}{2\pi}+\frac{15}{2}=14.\,980\,28
\]

\subsection{The sinh-Gordon model}

\label{asg}

The classical sinh-Gordon Lagrangian is%

\begin{equation}
\mathcal{L}^{SG}=\tfrac{1}{2}\partial_{\mu}\varphi\partial^{\mu}\varphi
+\frac{\alpha}{\beta^{2}}\left(  \cosh\beta\varphi-1\right)  =\tfrac{1}%
{2}\partial_{\mu}\varphi\partial^{\mu}\varphi-\frac{1}{2}\alpha\varphi
^{2}+\beta^{2}\alpha\frac{1}{24}\varphi^{4}+O\left(  \beta^{3}\right)
\allowbreak\label{Lsg}%
\end{equation}
and the field equation%
\[
\square\varphi(t,x)+\frac{\alpha}{\beta}\sinh\beta\varphi(t,x)=0.
\]
with%
\[
0<\mu=\frac{\beta^{2}}{8\pi+\beta^{2}}<1
\]
The model is super-renormalizable, therefore after introducing normal products
in (\ref{Lsg}) there are only two finite renormalization constants. The wave
function and the mass renormalization constants are given by \cite{KW,BK1}
\begin{equation*}
\langle\,0\,|\,\varphi(0)\,|\,p\,\rangle   =\sqrt{Z^{\varphi}}\,,~~~
\alpha   =m^{2}\frac{\pi\mu}{\sin\pi\mu}\,.
\end{equation*}
with \cite{KW}%
\[
Z^{\varphi}=(1-\mu)\frac{\frac{\pi}{2}\mu}{\sin\frac{\pi}{2}\mu}%
E(-\mu)\,,~~E(x)=\exp\left(  -\pi\int_{0}^{x}\frac{t}{\sin\pi t}dt\right)  \,.
\]
The \textbf{S-matrix} can be obtained by analytic continuation (from
$\beta\rightarrow i\beta$) of the sine-Gordon S-matrix which was derived in
\cite{KT,Za2}%
\[
S^{SG}(x)=\frac{\sinh\theta-i\sin\pi\mu}{\sinh\theta+i\sin\pi\mu}=-\exp\left(
-2\int_{0}^{\infty}\frac{dt}{t}\,\frac{\cosh\left(  \frac{1}{2}-\mu\right)
t}{\cosh\frac{1}{2}t}\sinh t\frac{\theta}{i\pi}\right)  .
\]
The minimal \textbf{sinh-Gordon form factor} is \cite{KW,BK2}
\begin{align}
F^{SG}(\theta)  &  =\exp\int_{0}^{\infty}\frac{dt}{t\sinh t}\,\left(
\frac{\cosh\left(  \frac{1}{2}-\mu\right)  t}{\cosh\frac{1}{2}t}-1\right)
\cosh t\left(  1-\frac{\theta}{i\pi}\right) \label{FSG}\\
&  =-i\sinh\tfrac{1}{2}\theta\,\xi\left(  \mu+\left(  1-\theta/\left(
i\pi\right)  \right)  \right)  \xi\left(  \mu-\left(  1-\theta/\left(
i\pi\right)  \right)  \right) \nonumber
\end{align}
where the meromorphic function\footnote{The function $E(x)$ was introduced in
\cite{KW} and also used in \cite{BFKZ} and \cite{BK}.}
\[
\xi(x)=\sqrt{\frac{1}{\cos\frac{1}{2}\pi x}E(x)}=%
{\textstyle\prod\limits_{k=0}^{\infty}}
\frac{\Gamma\left(  1+k-\frac{1}{2}x\right)  \Gamma\left(  \frac{1}{2}%
+k+\frac{1}{2}x\right)  }{\Gamma\left(  \frac{3}{2}+k-\frac{1}{2}x\right)
\Gamma\left(  1+k+\frac{1}{2}x\right)  }\frac{\Gamma\left(  \frac{3}%
{2}+k\right)  }{\Gamma\left(  \frac{1}{2}+k\right)  }%
\]
has been introduced, for more details see \ref{a5}. The 3-particle form factor
is \cite{KW,BK2}%
\begin{equation}
F^{SG}(\theta_{1},\theta_{2},\theta_{3})=-\sqrt{Z^{\varphi}}\frac{\sin\pi\mu
}{F\left(  i\pi\right)  }\frac{F(\theta_{12})F(\theta_{13})F(\theta_{23}%
)}{\cosh\frac{1}{2}\theta_{12}\,\cosh\frac{1}{2}\theta_{13}\,\cosh\frac{1}%
{2}\theta_{23}}\label{Fsg3}%
\end{equation}
where the normalization follows from the form factor equation (iii) and
(\ref{FSG})%
\begin{align*}
\operatorname*{Res}_{\theta_{12}=i\pi}F^{SG}(\theta_{1},\theta_{2},\theta
_{3})  &  =2i\,\left(  \mathbf{1}-S(\theta_{23})\right)  \sqrt{Z^{\varphi}}\\
F^{SG}(\theta+i\pi)F^{SG}(\theta)  &  =\frac{\sinh\theta}{\sinh\theta+i\sin
\pi\mu}\,.
\end{align*}

\subsubsection{The 4-point function}

From (\ref{Sigma4}) for $k_{i}=(k_{i}^{0},0)$ we get with $\underline{\varphi
}=\varphi\varphi\varphi\varphi$
\[
\tilde{\Xi}_{\underline{\varphi}}^{SG}(\underline{k})=\frac{-i}{32\pi m^{6}%
}\sum_{perm(k)}\frac{m}{-k_{1}^{0}+m-i\epsilon}\frac{m}{k_{4}^{0}+m-i\epsilon
}g^{SG}\left(  \frac{-1}{2m}\left(  k_{3}^{0}+k_{4}^{0}\right)  \right)
\]
where $g^{SG}(x)=g_{1}^{SG}(x)+g_{2}^{SG}(x)$ and%
\[
g_{i}^{SG}(x)=-\frac{1}{4}\int\frac{1}{\cosh\theta}\frac{1}{\cosh\theta
-x}I_{\underline{\varphi}i}^{SG}(0,\theta,-\theta,0)d\theta\,.
\]
From (\ref{I}) and (\ref{Fsg3}) we obtain
\begin{align*}
I_{\underline{\varphi}1}^{SG}(0,\theta,-\theta,0)  &  =\tfrac{1}{2}Z^{\varphi
}F^{SG}(0,\theta-i\pi_{+},-\theta-i\pi_{-})F^{SG}(-\theta+i\pi_{+},\theta
+i\pi_{-},0)+\left(  \epsilon\rightarrow-\epsilon\right) \\
&  =f^{SG}(\theta)I_{\underline{\varphi}}^{Z_{2}}(0,\theta,-\theta,0)
\end{align*}
where $I_{\underline{\varphi}}^{Z_{2}}$ as defined in (\ref{IZ2}). We have
introduced%
\[
f^{SG}(\theta)=-\frac{\left(  Z^{\varphi}\right)  ^{2}\sin^{2}\pi\mu}%
{F^{2}\left(  i\pi\right)  \left(  2i\right)  ^{2}}\left(  F_{0}(\theta
+i\pi)\right)  ^{4}F_{0}(2\theta)F_{0}(-2\theta)
\]
with%
\[
F_{0}(\theta)=F^{SG}(\theta)/\left(  -i\sinh\tfrac{1}{2}\theta\right)
=\,\xi\left(  \mu+\left(  1-\theta/\left(  i\pi\right)  \right)  \right)
\xi\left(  \mu-\left(  1-\theta/\left(  i\pi\right)  \right)  \right)  \,.
\]
Therefore as in (\ref{gZ2}) and (\ref{IZ2}) we obtain%
\begin{align*}
g_{1}^{SG}(x)  &  =-\frac{1}{4}\int\frac{1}{\cosh\theta}\frac{1}{\cosh
\theta-x}I_{1}^{SG}(0,\theta,-\theta,0)d\theta\\
&  =-\int_{-\infty}^{\infty}\left(  f^{SG}(\theta)\frac{\coth^{4}\tfrac{1}%
{2}\theta\,\tanh^{2}\theta}{\cosh\theta}\frac{1}{\cosh\theta-x}-f^{SG}%
(0)\frac{16}{\theta^{2}}\frac{1}{1-x}\right)  d\theta.
\end{align*}
The functions $g_{1}^{SG}(x)$ for $\mu=0.3$ and $\mu=0.5$ are plotted in Fig.
\ref{sgg3} and \ref{sgg5}.

\label{2}

The contribution from $I_{2}$ follows from (\ref{g2}) as%
\[
g_{2}^{SG}\left(  x\right)  =-16\pi i\frac{dS(\theta)}{d\theta}\sqrt
{Z^{\varphi}}\frac{1}{1-x-i\epsilon}=\frac{-32\pi}{\sin\pi\mu}\sqrt
{Z^{\varphi}}\frac{1}{1-x}.
\]

\subsubsection{Properties of $\xi(x)$}

\label{a5}

Representations%
\begin{align*}
\xi(x)  &  =\exp\frac{1}{2}\int_{0}^{\infty}\frac{dt}{t\sinh t}\,\left(
\frac{\cosh\left(  \frac{1}{2}-x\right)  t}{\cosh\frac{1}{2}t}-1\right) \\
&  =\exp\left\{  \frac{1}{2}\left(  \frac{i}{\pi}\left(  \operatorname*{Li}%
\nolimits_{2}(e^{ix\pi})-\operatorname*{Li}\nolimits_{2}(-e^{ix\pi})\right)
\right.  \right. \\
&  -\left.  \left.  x\ln\left(  1-e^{ix\pi}\right)  -\left(  1-x\right)
\ln\left(  1+e^{ix\pi}\right)  +\ln2+\tfrac{1}{2}i\pi\left(  x-\tfrac{1}%
{2}\right)  \right)  \right\} \\
&  =\frac{1}{\sqrt{\pi}}\frac{G\left(  1+\frac{1}{2}x\right)  G\left(
\frac{3}{2}-\frac{1}{2}x\right)  }{G\left(  1-\frac{1}{2}x\right)  G\left(
\frac{1}{2}+\frac{1}{2}x\right)  }%
\end{align*}
where $G$ is Barnes G-function and $\operatorname*{Li}\nolimits_{2}(x)$ the
dilogarithm\footnote{In Mathematica: $\operatorname*{Li}\nolimits_{2}%
(x)=\operatorname*{PolyLog}[2,x]$.}. The function $\xi(x)$ is meromorphic and
satisfies%
\begin{equation*}
\xi(1-x)=\xi(x)\,,~~~
\xi(x)\xi(-x)\cos\tfrac{1}{2}\pi x=\sqrt{E(x)E(-x)}=1
\end{equation*}
which imply the form factor equation (i) $F^{SG}(x)=F^{SG}(-x)S^{SG}(x)$.

\subsection{The $Z_{3}$-model}

\label{aZ3}

The two-particle S-matrix for the $Z_{N}$-Ising model has been proposed by
K\"{o}berle and Swieca \cite{KS}. The scattering of two particles of type $1$
is given by%
\begin{equation}
S(\theta)=\frac{\sinh\frac{1}{2}(\theta+\frac{2\pi i}{N})}{\sinh\frac{1}%
{2}(\theta-\frac{2\pi i}{N})}\,.
\end{equation}
This S-matrix is consistent with the picture that the bound state of $N-1$
particles of type $1$ is the anti-particle of $1$. The form factors of the
$Z_{N}$-model have been proposed in \cite{K3,BFK}. The minimal solution of
Watson's and the crossing equations
\[
F(\theta)=F(-\theta)S(\theta)\,,~~F(i\pi-\theta)=F(i\pi+\theta)
\]
for the $Z_{3}$ model is%
\begin{align*}
F^{Z3}(i\pi x)  &  =c_{1}\sin\tfrac{1}{2}\pi x\,\exp\int_{0}^{\infty}\left(
\frac{\sinh\frac{1}{3}t}{t\sinh^{2}t}\left(  1-\cosh t\left(  1-x\right)
\right)  \right)  dt\\
&  =\pi^{1/3}\sin\tfrac{1}{2}\pi x\,\frac{G\left(  \frac{1}{3}+\frac{1}%
{2}x\right)  G\left(  \frac{4}{3}-\frac{1}{2}x\right)  }{G\left(  \frac{2}%
{3}+\frac{1}{2}x\right)  G\left(  \frac{5}{3}-\frac{1}{2}x\right)  }%
\end{align*}
where $G(x)$ is Barnes G-function.

The form factor of the order parameter $\sigma_{1}(x)$ and two particles of
type $2$ is given by (\ref{F22}) \cite{K3,BFK} where $c_{2}=-\frac{1}{2}%
\sqrt{2}\sqrt[4]{3}\pi^{-\frac{1}{3}}G\left(  \frac{4}{3}\right)  /G\left(
\frac{2}{3}\right)  $ is determined by the form factor equation (iv)
$\operatorname*{Res}_{\theta_{12}=\frac{2}{3}i\pi}F_{22}^{\sigma_{1}%
}(\underline{\theta})=\sqrt{2}F_{1}^{\sigma_{1}}\Gamma=\sqrt{2}\Gamma$. The
intertwiner $\Gamma$ defined by is defined by%
\[
i\operatorname*{Res}_{\theta=\frac{2}{3}i\pi}\frac{\sinh\frac{1}{2}(i\pi
x+\frac{2\pi i}{3})}{\sinh\frac{1}{2}(i\pi x-\frac{2\pi i}{3})}=-\sqrt
{3}=\Gamma_{2}^{11}\Gamma_{11}^{2}\,,~~\Gamma_{2}^{11}=\Gamma_{11}^{2}%
=\Gamma=i3^{\frac{1}{4}}.
\]
The form factor of $\sigma_{1}(x)$ for the 3 particles of type $112$ is given
by (\ref{F112}) \cite{K3,BFK}, where $c_{3}=\sqrt{3}\pi^{\frac{2}{3}}\left(
G\left(  \frac{4}{3}\right)  \right)  ^{2}/\left(  G\left(  \frac{2}%
{3}\right)  \right)  ^{2}$ is determined by $\operatorname*{Res}_{\theta
_{12}=\frac{2}{3}i\pi}F_{112}^{\sigma_{1}}(\underline{\theta})=\sqrt{2}%
F_{22}^{\sigma_{1}}(\underline{\theta})\Gamma$. The minimal form factor of the
particles $1$ and $2$%
\[
F_{(12)}^{\min}(i\pi x)   =c\exp\int_{0}^{\infty}\frac{dt}{t\sinh^{2}t}%
\sinh\frac{2}{3}t\,\left(  1-\cosh t\left(  1-x\right)  \right)
 =\frac{G\left(  \frac{1}{6}+\frac{1}{2}x\right)  G\left(  \frac{7}{6}%
-\frac{1}{2}x\right)  }{G\left(  \frac{5}{6}+\frac{1}{2}x\right)  G\left(
\frac{11}{6}-\frac{1}{2}x\right)  }%
\]
satisfies $F_{(12)}^{\min}(\theta)=F_{(12)}^{\min}(-\theta)S_{(12)}(\theta)$,
where $S_{(12)}(\theta)=-\frac{\sinh\frac{1}{2}\left(  \theta+\frac{1}{3}%
i\pi\right)  }{\sinh\frac{1}{2}\left(  \theta-\frac{1}{3}i\pi\right)  }$ is
the the S-matrix for the particles $1$ and $2$.

\subsubsection{The 3-point function}

The three point function Green's function (see also \cite{CDGJM})
\newline$\tau_{\sigma_{1}\sigma
_{1}\sigma_{1}}(\underline{x})=\langle\,0\,|\,T\sigma_{1}(x_{1})\sigma
_{1}(x_{2})\sigma_{1}(x_{3})\,|\,0\,\rangle$ of the field $\sigma_{1}(x)$ is
different from zero, because $\sigma_{1}\sigma_{1}\sigma_{1}$ is in the vacuum
sector. We have the contributions
\begin{multline*}
\tilde{\Xi}_{\sigma_{1}\sigma_{1}\sigma_{1}}(\underline{k})=\sum_{\pi\in
S_{3}}\left(  \tilde{\Xi}_{\sigma_{1}\sigma_{1}\sigma_{1}}^{11}(k_{\pi
1},k_{\pi2},k_{\pi3})+\tilde{\Xi}_{\sigma_{1}\sigma_{1}\sigma_{1}}^{12}%
(k_{\pi1},k_{\pi2},k_{\pi3})\right.  \\
+\left.  \tilde{\Xi}_{\sigma_{1}\sigma_{1}\sigma_{1}}^{\bar{2}\bar{1}}%
(k_{\pi1},k_{\pi2},k_{\pi3})\right)  \,.
\end{multline*}
For the permutation $\pi=(1,2,3)$ and the intermediate states\newline%
$\left\langle 0|\sigma_{1}(0)|p_{1}\rangle\langle p_{1}|\sigma_{1}(0)|\bar
{p}_{2}\rangle\langle\bar{p}_{2}|\,\sigma_{1}(0)|0\right\rangle $ the
$\tilde{\Xi}$-function is as in (\ref{Xi11})%
\begin{align*}
\tilde{\Xi}_{\sigma_{1}\sigma_{1}\sigma_{1}}^{11}(k_{1},k_{2},k_{3}) &
=\int_{p_{1}}\int_{p_{2}}\left\langle 0|\sigma_{1}(0)|p_{1}\rangle\langle
p_{1}|\sigma_{1}(0)|\bar{p}_{2}\rangle\langle\bar{p}_{2}|\,\sigma
_{1}(0)|0\right\rangle \\
&  \times2\pi\delta\left(  p_{1}\right)  2\pi\delta\left(  \bar{p}_{2}\right)
\frac{-i}{k_{2}^{0}+k_{3}^{0}+\omega_{1}-i\epsilon}\frac{-i}{k_{3}^{0}%
+\omega_{2}-i\epsilon}\\
&  =\frac{-1}{4m^{4}}\frac{m}{-k_{1}^{0}+m-i\epsilon}\frac{m}{k_{3}%
^{0}+m-i\epsilon}F_{22}^{\sigma_{1}}(i\pi,0)
\end{align*}
For the permutation $\pi=(1,2,3)$ and the intermediate states\newline%
$\left\langle 0|\sigma_{1}(0)|p_{1}\rangle\langle p_{1}|\sigma_{1}%
(0)|p_{2},p_{3}\rangle\langle p_{3},p_{2}|\,\sigma_{1}(0)|0\right\rangle $ the
$\tilde{\Xi}$-function is as in (\ref{Xi12})%
\begin{align*}
\tilde{\Xi}_{\sigma_{1}\sigma_{1}\sigma_{1}}^{12}(k_{1},k_{2},k_{3})  &
=\frac{1}{2!}\int_{p_{1}}\int_{p_{2}}\int_{p_{3}}\left\langle 0|\sigma
_{1}(0)|p_{1}\rangle\langle p_{1}|\sigma_{1}(0)|p_{2},p_{3}\rangle\langle
p_{3},p_{2}|\,\sigma_{1}(0)|0\right\rangle \\
&  \times2\pi\delta\left(  p_{1}\right)  2\pi\delta\left(  p_{2}+p_{3}\right)
\frac{-i}{k_{2}^{0}+k_{3}^{0}+\omega_{1}-i\epsilon}\frac{-i}{k_{3}^{0}%
+\omega_{2}+\omega_{3}-i\epsilon}\\
&  =-\frac{1}{64\pi m^{4}}\frac{m}{-k_{1}^{0}+m-i\epsilon}\int d\theta\frac
{m}{\omega}\frac{2m}{k_{3}^{0}+2\omega-i\epsilon}I_{\sigma_{1}\sigma_{1}%
\sigma_{1}}^{12}(\theta)\\
I_{\sigma_{1}\sigma_{1}\sigma_{1}}^{12}(\theta)  &  =F_{211}^{\sigma_{1}}%
(i\pi,\theta,-\theta)F_{22}^{\sigma_{1}}(-\theta+i\pi,\theta+i\pi)
\end{align*}
where the crossing relation%
\[
\langle p_{1}|\sigma_{1}(0)|p_{2},p_{3}\rangle=F_{211}^{\sigma_{1}}(\theta
_{1}+i\pi,\theta_{2},\theta_{3})+\delta_{\theta_{1}\theta_{2}}+\delta
_{\theta_{1}\theta_{3}}S(\theta_{23})
\]
has been used. The $\delta$-terms do not contribute because $F_{22}%
^{\sigma_{1}}(0,0)=0$. Inserting the form factor functions we get
\begin{multline*}
\tilde{\Xi}_{\sigma_{1}\sigma_{1}\sigma_{1}}^{12}(k_{1},k_{2},k_{3})=-\frac
{1}{64\pi m^{4}}\frac{m}{-k_{1}^{0}+m-i\epsilon}h^{Z3}\left(  -\frac{k_{3}%
^{0}}{2m}+i\epsilon\right) \\
h^{Z3}\left(  x\right)  =\int_{-\infty}^{\infty}d\theta\frac{1}{\cosh\theta
}\frac{1}{\cosh\theta-x}\frac{F(2\theta)F(-2\theta)F_{(12)}^{\min}(i\pi
+\theta)F_{(12)}^{\min}(i\pi-\theta)}{\left(  \sinh(\theta-\frac{1}{3}%
i\pi)\,\sinh(\theta+\frac{1}{3}i\pi)\,\sinh\frac{1}{2}\theta\right)  ^{2}}.
\end{multline*}
For the intermediate states $\left\langle 0|\sigma_{1}(0)|\bar{p}_{1},\bar
{p}_{2}\rangle\langle\bar{p}_{2},\bar{p}_{1}|\sigma_{1}(0)|\bar{p}_{3}%
\rangle\langle\bar{p}_{3}|\,\sigma_{1}(0)|0\right\rangle $, where $|\bar
{p}\rangle$ is a particle state of type 2 which is the anti-particle of 1, the
$\tilde{\Xi}$-function is as in (\ref{Xi21})%
\begin{align*}
\tilde{\Xi}_{\sigma_{1}\sigma_{1}\sigma_{1}}^{\bar{2}\bar{1}}(\underline{k})
&  =\frac{1}{2!}\int_{p_{1}}\int_{p_{2}}\int_{p_{3}}\left\langle 0|\sigma
_{1}(0)|\bar{p}_{1},\bar{p}_{2}\rangle\langle\bar{p}_{2},\bar{p}_{1}%
|\sigma_{1}(0)|\bar{p}_{3}\rangle\langle\bar{p}_{3}|\,\sigma_{1}%
(0)|0\right\rangle \\
&  \times2\pi\delta\left(  p_{1}\right)  2\pi\delta\left(  p_{2}+p_{3}\right)
\frac{-i}{\pi k_{2}^{0}+\pi k_{3}^{0}+\omega_{1}+\omega_{2}-i\epsilon}%
\frac{-i}{\pi k_{3}^{0}+\omega_{3}-i\epsilon}\\
&  =-\frac{1}{64\pi m^{4}}\frac{m}{k_{2}^{0}+k_{3}^{0}+m-i\epsilon}\int
d\theta\frac{m}{\omega}\frac{2m}{-k_{1}^{0}+2\omega-i\epsilon}I_{\sigma
_{1}\sigma_{1}\sigma_{1}}^{\bar{2}\bar{1}}(\theta)\\
I_{\sigma_{1}\sigma_{1}\sigma_{1}}^{\bar{2}\bar{1}}(\theta)  &  =F_{22}%
^{\sigma_{1}}(\theta,-\theta)F_{112}^{\sigma_{1}}(-\theta+i\pi,\theta+i\pi,0)
\end{align*}
where (\ref{F22}), (\ref{F112}) and the crossing relation%
\[
\langle\bar{p}_{2},\bar{p}_{1}|\sigma_{1}(0)|\bar{p}_{3}\rangle=\left(
F_{112}^{\sigma_{1}}(\theta_{2}+i\pi,\theta_{1}+i\pi,\theta_{3})+\delta
_{\theta_{1}\theta_{3}}+S(\theta_{21})\delta_{\theta_{2}\theta_{3}}\right)
\]
have been used. Again the $\delta$-terms do not contribute because
$F_{22}^{\sigma_{1}}(0,0)=0$. Inserting the form factor functions we get%

\[
\tilde{\Xi}_{\sigma_{1}\sigma_{1}\sigma_{1}}^{\bar{2}\bar{1}}(k_{1}%
,k_{2},k_{3})=-\frac{1}{64\pi m^{4}}\frac{m}{k_{3}^{0}+m-i\epsilon}%
h^{Z3}\left(  \frac{k_{1}^{0}}{2m}+i\epsilon\right)
\]
and as in (\ref{Xi3})
\begin{align*}
\tilde{\Xi}_{\sigma_{1}\sigma_{1}\sigma_{1}}(k_{1},k_{2},k_{3})  &
=const.\sum_{perm(k)}\frac{m}{-k_{1}^{0}+m-i\epsilon}\frac{m}{k_{3}%
^{0}+m-i\epsilon}\\
& +const.^{\prime}\sum_{perm(k)}\frac{m}{-k_{1}^{0}+m-i\epsilon}%
h^{Z(3)}\left(  -\frac{k_{3}^{0}}{2m}+i\epsilon\right)  +\left(
k_{i}\rightarrow-k_{i}\right)  \,.
\end{align*}
For $h^{Z3}(x)$ see Fig. \ref{hz3}.

\subsubsection{The 4-point function}

We consider the Green's function\newline$\tau_{\sigma_{1}\sigma_{2}\sigma
_{1}\sigma_{2}}(\underline{x})=\langle\,0\,|\,T\sigma_{1}(x_{1})\sigma
_{2}(x_{2})\sigma_{1}(x_{3})\sigma_{2}(x_{4})\,|\,0\,\rangle$ and calculate
for $\sigma_{3}=\sigma_{1}$ and $\sigma_{4}=\sigma_{2}$%
\[
\tilde{\Xi}_{\sigma_{1}\sigma_{2}\sigma_{3}\sigma_{4}}(k_{1},k_{2},k_{3}%
,k_{4})=\sum_{\pi\in S_{4}}\tilde{\Xi}_{\sigma_{\pi1}\sigma_{\pi2}\sigma
_{\pi3}\sigma_{\pi4}}^{121}(k_{\pi1},k_{\pi2},k_{\pi3},k_{\pi4})\,
\]
where as in (\ref{I123}) the result is expressed by 3 terms $I_{1},I_{2}%
,I_{3}$. The $I_{3}$ term again contributes to the disconnected part and the
$I_{2}$ is as in (\ref{gsg2}) given by $g_{2}=const./(1-x)$. We calculate here
the more interesting contribution from $I_{1}$. For the various permutations
we have to calculate as in (\ref{Xigen0}):

\label{1}\bigskip

$\mathbf{I)}$ for the permutation $\pi=(1,2,3,4)$ and the intermediate
states\newline$\left\langle 0|\sigma_{1}(0)|p_{1}\rangle\allowbreak\langle
p_{1}|\sigma_{2}(0)|\bar{p}_{2},p_{3}\rangle\allowbreak\langle p_{3},\bar
{p}_{2}|\,\sigma_{1}(0)|p_{4}\rangle\allowbreak\langle p_{4}|\,\sigma
_{2}(0)|0\right\rangle $ we obtain%
\[
\tilde{\Xi}_{\sigma_{1}\sigma_{2}\sigma_{1}\sigma_{2}}^{121}(k_{1},k_{2}%
,k_{3},k_{4})=-\frac{1}{32}\frac{i}{m^{6}\pi}\frac{m}{-k_{1}^{0}+m-i\epsilon
}\frac{m}{k_{4}^{0}+m-i\epsilon}g_{I}^{Z3}\left(  -\frac{k_{3}^{0}+k_{4}^{0}%
}{2m}+i\epsilon\right)
\]%
\begin{gather*}
g_{I}^{Z3}(x)=\frac{-1}{2}\int d\theta\frac{1}{\cosh\theta}\frac{1}%
{\cosh\theta-x}I_{\sigma_{1}\sigma_{2}\sigma_{1}\sigma_{2}}^{121}%
(0,\theta,-\theta,0)\\
I_{\sigma_{1}\sigma_{2}\sigma_{1}\sigma_{2}}^{121}(\theta_{1},\theta
_{2},\theta_{3},\theta_{4})=\langle p_{1}|\sigma_{2}(0)|\bar{p}_{2}%
,p_{3}\rangle\langle p_{3},\bar{p}_{2}|\,\sigma_{1}(0)|p_{4}\rangle_{1}%
\end{gather*}
where $\langle\dots\rangle_{1}$ means that we only take into account the term
from $I_{1}$ and as in (\ref{I})
\begin{multline*}
I_{\sigma_{1}\sigma_{2}\sigma_{1}\sigma_{2}}^{121}(0,\theta,-\theta,0)\\
=\tfrac{1}{2}F_{221}^{\sigma_{2}}(0,\theta-i\pi_{+},-\theta-i\pi_{-}%
)F_{211}^{\sigma_{1}}(-\theta+i\pi_{+},\theta+i\pi_{-},0)+\left(  \pi
_{+}\leftrightarrow\pi_{-}\right) \\
=\tfrac{1}{2}\frac{c_{3}^{2}\left(  F^{Z3}\left(  i\pi-\theta\right)
F_{12}^{\min}\left(  i\pi-\theta\right)  \right)  ^{2}F_{12}^{\min}\left(
2\theta\right)  F_{12}^{\min}\left(  -2\theta\right)  }{\left(  \sinh\frac
{1}{2}\left(  \theta-\frac{1}{3}i\pi\right)  \sinh\frac{1}{2}\left(
\theta+\frac{1}{3}i\pi\right)  \cosh\theta\right)  ^{2}}+\left(
\epsilon\rightarrow-\epsilon\right)
\end{multline*}
where (\ref{F112}) and charge conjugation invariance $F_{221}^{\sigma_{2}%
}\left(  \underline{\theta}\right)  =F_{112}^{\sigma_{1}}\left(
\underline{\theta}\right)  $ have been used. For $g_{I}^{Z3}(x)$ see Fig.
\ref{gz31}. For the permutations\newline$\pi=(3,2,1,4),~(1,4,3,2),~(3,4,1,2)$
the result is, similarly, expressed by $g_{I}^{Z3}(x)$.

\label{here}$\bigskip$

$\mathbf{II)}$ for $\pi=(1,3,2,4)$ and the intermediate states\newline%
$\left\langle 0|\sigma_{1}(0)|p_{1}\rangle\langle p_{1}|\sigma_{1}%
(0)|p_{2},p_{3}\rangle\langle p_{3},p_{2}|\,\sigma_{2}(0)|p_{4}\rangle\langle
p_{4}|\,\sigma_{2}(0)|0\right\rangle $ we obtain%
\[
\tilde{\Xi}_{\sigma_{1}\sigma_{1}\sigma_{2}\sigma_{2}}^{121}(k_{1},k_{3}%
,k_{2},k_{4})=-\frac{1}{32}\frac{i}{m^{6}\pi}\frac{m}{-k_{1}^{0}+m-i\epsilon
}\frac{m}{k_{4}^{0}+m-i\epsilon}g_{II}^{Z3}\left(  -\frac{k_{2}^{0}+k_{4}^{0}%
}{2m}+i\epsilon\right)
\]%
\begin{gather*}
g_{II}^{Z3}(x)=\frac{-1}{4}\int d\theta\frac{1}{\cosh\theta}\frac{1}%
{\cosh\theta-x}I_{\sigma_{1}\sigma_{1}\sigma_{2}\sigma_{2}}^{121}%
(0,\theta,-\theta,0)\\
I_{\sigma_{1}\sigma_{1}\sigma_{2}\sigma_{2}}^{121}(\theta_{1},\theta
_{2},\theta_{3},\theta_{4})=\langle p_{1}|\sigma_{1}(0)|p_{2},p_{3}%
\rangle\langle p_{3},p_{2}|\,\sigma_{2}(0)|p_{4}\rangle_{1}%
\end{gather*}
where $\langle\dots\rangle_{1}$ means that we only consider the term from
$I_{1}$ and as in (\ref{I})%
\begin{align*}
&  I_{\sigma_{1}\sigma_{1}\sigma_{2}\sigma_{2}}^{121}(0,\theta,-\theta,0)\\
&  =\tfrac{1}{2}F_{211}^{\sigma_{1}}(0,\theta-i\pi_{+},-\theta-i\pi
_{-})F_{221}^{\sigma_{2}}(-\theta+i\pi_{+},\theta+i\pi_{-},0)+\left(  \pi
_{+}\leftrightarrow\pi_{-}\right)  .
\end{align*}
For $g_{II}^{Z3}(x)$ see Fig. \ref{gz32}. For the permutations \newline%
$\pi=(3,1,2,4),~(1,3,4,2),~(3,1,4,2)$ the result is, similarly, expressed by
$g_{II}^{Z3}(x)$.$\bigskip$

$\mathbf{III)}$ for $\pi=(1,2,4,3)$ and the intermediate states\newline%
$\left\langle 0|\sigma_{1}(0)|p_{1}\rangle\langle p_{1}|\sigma_{2}(0)|\bar
{p}_{2},p_{3}\rangle\langle p_{3},\bar{p}_{2}|\,\sigma_{2}(0)|\bar{p}%
_{4}\rangle\langle\bar{p}_{4}|\,\sigma_{1}(0)|0\right\rangle $ we obtain%
\[
\tilde{\Xi}_{\sigma_{1}\sigma_{2}\sigma_{2}\sigma_{1}}^{121}(k_{1},k_{2}%
,k_{4},k_{3})=-\frac{1}{32}\frac{i}{m^{6}\pi}\frac{m}{-k_{1}^{0}+m-i\epsilon
}\frac{m}{k_{3}^{0}+m-i\epsilon}g_{III}^{Z3}\left(  -\frac{k_{4}^{0}+k_{3}%
^{0}}{2m}+i\epsilon\right)
\]%
\begin{gather*}
g_{III}^{Z3}(x)=\frac{-1}{4}\int d\theta\frac{1}{\cosh\theta}\frac{1}%
{\cosh\theta-x}I_{\sigma_{1}\sigma_{2}\sigma_{2}\sigma_{1}}^{121}%
(0,\theta,-\theta,0)\\
I_{\sigma_{1}\sigma_{2}\sigma_{2}\sigma_{1}}^{121}(\theta_{1},\theta
_{2},\theta_{3},\theta_{4})=\langle p_{1}|\sigma_{2}(0)|\bar{p}_{2}%
,p_{3}\rangle\langle p_{3},\bar{p}_{2}|\,\sigma_{2}(0)|\bar{p}_{4}\rangle_{1}%
\end{gather*}
where $\langle\dots\rangle_{1}$ means that we only consider the term from
$I_{1}$ and as in (\ref{I})%
\begin{multline*}
I_{\sigma_{1}\sigma_{2}\sigma_{2}\sigma_{1}}^{121}(0,\theta,-\theta,0)\\
=\tfrac{1}{2}F_{221}^{\sigma_{2}}(0,\theta-i\pi_{+},-\theta-i\pi_{-}%
)F_{212}^{\sigma_{2}}(-\theta+i\pi_{+},\theta+i\pi_{-},0)+\left(  \pi
_{+}\leftrightarrow\pi_{-}\right)  .
\end{multline*}
It turns out that $I_{\sigma_{1}\sigma_{2}\sigma_{2}\sigma_{1}}^{121}%
(0,\theta,-\theta,0)=I_{\sigma_{1}\sigma_{2}\sigma_{1}\sigma_{2}}%
^{121}(0,\theta,-\theta,0)$ which follows from charge conjugation invariance
$F_{221}^{\sigma_{2}}\left(  \theta_{1},\theta_{2},\theta_{3}\right)
=F_{112}^{\sigma_{1}}\left(  \theta_{1},\theta_{2},\theta_{3}\right)  $,
therefore
\[
g_{III}^{Z3}(x)=g_{I}^{Z3}(x)\,.
\]
For the permutations $\pi=(3,2,4,1),~(1,4,2,3),~(3,4,2,1)$ the result is,
similarly, expressed by $g_{I}^{Z3}(x)$.


\end{document}